\documentclass[
showkeys,	
reprint,	
raggedbottom,	
flushbottom,
amsmath,	
amssymb,	
amsfonts,	
aps,		
pre,%
groupedaddress,
superscriptaddress,
a4paper,%
nofootinbib,
floatfix,	%
preprintnumbers, 
]{revtex4-2}

\usepackage{amsthm,mathtools}

\usepackage{subfigure}

\usepackage{multirow}
\usepackage{physics}
\usepackage{graphics,graphicx} 
\usepackage{wasysym}

\usepackage{xcolor, color}
\definecolor{BlueGreen}{RGB}{0, 154, 166}      
\definecolor{VioletRed}{RGB}{215, 31, 133}     
\definecolor{Purple}{RGB}{182, 52, 187} 	
\definecolor{Turquoise}{RGB}{0, 255, 239} 	   
\definecolor{amber}{rgb}{1.0, 0.75, 0.0}
\definecolor{uscgold}{rgb}{1.0, 0.8, 0.0}
\definecolor{uclagold}{rgb}{1.0, 0.7, 0.0}
\definecolor{gold(metallic)}{rgb}{0.83, 0.69, 0.22}
\definecolor{gold(web)(golden)}{rgb}{1.0, 0.84, 0.0}
\definecolor{goldenpoppy}{rgb}{0.99, 0.76, 0.0}
\definecolor{goldenyellow}{rgb}{1.0, 0.87, 0.0}
\definecolor{goldenrod}{rgb}{0.85, 0.65, 0.13}	   

\usepackage[
			]{hyperref}

\usepackage{natbib}
\hypersetup{
colorlinks=true,
linkcolor={Purple},
filecolor=magenta,      
urlcolor={BlueGreen},
citecolor=cyan,
anchorcolor=orange,
linktocpage=true,
breaklinks,
breaklinks=true,
pdftitle={Classical and quantum chaos in bean- and peanut-shaped billiards},
pdfauthor={Pranaya Pratik Das},
final
}

\usepackage{booktabs}

\usepackage{float}
\usepackage{tabularray}

\usepackage{soul, pifont}

\usepackage[capitalise]{cleveref}
\crefname{figure}{FIG.}{FIGS.}
\Crefname{figure}{FIG.}{FIGS.} 
\makeatletter
\AtBeginDocument
{
	\def\ltx@label#1{\cref@label{#1}}
	\def\label@in@display@noarg#1{\cref@old@label@in@display{#1}}
	\def\label@in@mmeasure@noarg#1{%
		\begingroup%
		\measuring@false%
		\cref@old@label@in@display{#1}
		\endgroup}%
} %
\makeatother

\usepackage{enumitem}

\usepackage{orcidlink}

\begin{document}
\title{Classical and quantum chaos in bean- and peanut-shaped billiards}

\author{Pranaya Pratik Das\,\orcidlink{0000-0002-6025-7719}}
\email{pranaya.phy@outlook.com}
\affiliation{Department of Physics and Astronomy, National Institute of Technology Rourkela, Odisha, India-769008}
\homepage[Homepage: \phantomsection]{https://ppdws.github.io/pranayapratikdas/}
\author{Tanmayee Patra\,\orcidlink{0009-0007-5555-0068}}%
\email{519ph1011@nitrkl.ac.in}
\affiliation{%
	Department of Physics and Astronomy, National Institute of Technology Rourkela, Odisha, India-769008}
\author{Biplab Ganguli\,\orcidlink{0000-0003-2583-5752}}%
\email{biplabg@nitrkl.ac.in}
\affiliation{%
	Department of Physics and Astronomy, National Institute of Technology Rourkela, Odisha, India-769008}%

\date{\today}

\begin{abstract}
	The geometry of a billiard boundary fundamentally governs its dynamics, ranging from integrable to mixed and fully chaotic regimes. Bean- and peanut-shaped billiards have varying curvature with both focusing and defocusing walls without a neutral segments. Particle dynamics inside these billiards show a strong correlation between classical and quantum dynamics in the chaotic regime also. This fundamental observation comes from our study of classical tools like Lyapunov exponent, Poincaré sections, flow trajectories in phase space and quantum tools that includes both statistical and dynamical measures. Statistical indicators include nearest-neighbour spacing distributions, level-spacing ratios, and the spectral staircase function, while dynamical measures include out-of-time-order correlators and spectral complexity. The dynamics in both of these billiard systems also exhibit eigenfunction scarring, an unexpected phenomenon observed in chaotic systems. Overall, our results provide a unified perspective on billiard systems with non-uniform curvature.
	
\end{abstract}


\maketitle
\pagestyle{plain}

\section{\label{sec1}Introduction}

\begin{figure}[hbt!]
	\centering
	\includegraphics[width=\linewidth]{"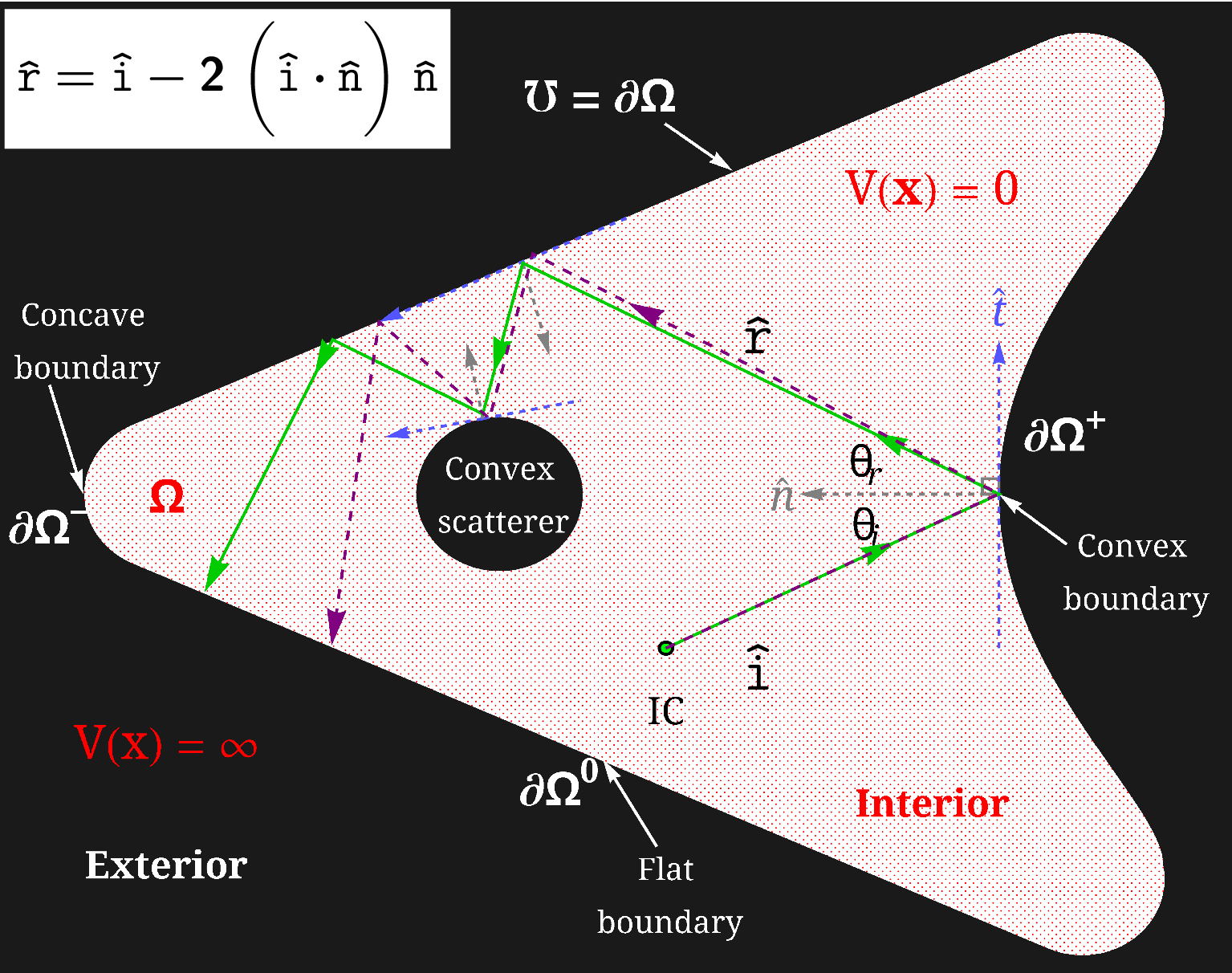"}
	\caption{A schematic representation of two nearby trajectory traced by a particle in a billiard domain ($ \varOmega $) with boundary $ \mho (= \partial{\varOmega}) $. Here,  $\hat{i}$, $\hat{r}$, and $\hat{n}$ are the incident vector, reflection vector, and normal vector at the point of collision, respectively. $\theta_{i}$ and $\theta_{r}$ are angle of incidence and angle of reflection, respectively and they follow the relation $\theta_{i}=\theta_{r}$ for every collision with the boundary.}
	\label{fig1}
\end{figure}

The exploration of nonlinear dynamics and chaos within Hamiltonian systems offers a multitude of untapped territories and fascinating subjects in both classical and quantum mechanics. Given that quantum mechanics provides a more fundamental description of the natural world, understanding how quantum systems behave when their classical counterparts are chaotic addresses deep questions about the emergence of statistical behaviour, thermalisation, and the quantum-classical boundary\cite{backer2003eigenfunctions, MichaelBerry1989, rspa.1987.0109}. While chaotic systems are broadly classified into dissipative and conservative categories, depending on whether phase-space volume contracts or is preserved during evolution, this work focuses on the conservative regime. In this scenario, billiard systems provide significant advantage because of their spatial confinement, which leads to specular ($=$ mirrorlike) reflections\footnote{Specular reflections refer to reflections in which particle reflects off the boundary with no change in the tangential component of momentum, and instantaneous reversal of the momentum component normal to the boundary.} \cite{Leonel2021, Sinai04, chernov2006chaotic}, as illustrated in \cref{fig1}. Billiards capture all the complexity of Hamiltonian dynamics, from integrable to chaotic, without the complexities of solving equations of motion.

In a billiard problem, a point particle moves freely without friction in a two-dimensional enclosed domain $\varOmega$. Between the elastic collisions at the boundary $ \mho $ ($ =\dd{\varOmega} $), the particle travels in straight lines with constant velocity\cite{Korsch2002, PhysRevLett.132.157101}. The billiards with static boundaries can be classified into, at least, three types: ($ i $) integrable billiards (e.g. Circular billiard \& Elliptical billiard); ($ ii $) chaotic billiards (e.g. Sinai billiard and the Bunimovich stadium) and; ($ iii $) mixed (co-existence of regular and chaotic dynamics) billiards. In the last case, the phase space shows a mixed behaviour with stable islands existing within the chaotic sea. The present work introduces two models that fall into the category of mixed billiards.

To illustrate the dynamics of a particle inside a billiard, we assume the Hamiltonian ($ \mathcal{H} $) of the particle of mass $ m $ which travels freely inside the billiard boundary $ \mho $  without friction is given by
\begin{equation}\label{eq1}
	\mathcal{H}(\mathbf{p}, \mathbf{x})=\dfrac{1}{2m}(\mathbf{p}^{2})+V(\mathbf{x}),
\end{equation} 
where,
\begin{equation}\label{eq2}
	V(\mathbf{x})={\begin{cases}
			0 & \text{If}\; \mathbf{x} \in \varOmega, \\
			\infty &  \text{If}\; \mathbf{x} \notin \varOmega.
	\end{cases}}
\end{equation}
Conventionally the mass and magnitude of momentum for the particle are set to $ m = \abs{p} = 1 $. The nature of this domain $\varOmega$ ensures the reflection dynamics in the billiard. The kinetic energy term in the system's Hamiltonian guarantees that the particle moves in a straight line between collisions, maintaining constant energy throughout its motion. Because of the inherently simplistic structure of this Hamiltonian, the equations that describe the particle's trajectory, known as the Hamilton-Jacobi equations, are equivalent to the geodesic equations on a manifold (i.e., the shortest paths between points in this domain, further emphasising the deterministic yet potentially chaotic nature of the system). It is to be noted that for a particle in billiard to display chaotic behaviour in both classical and quantum dynamics, it is essential for it to strictly adhere to the following properties\cite{CASATI1999293}: sensitive to \textit{Initial Conditions} (ICs), topologically mixing, ergodic, and positive Lyapunov exponents.

Chaotic billiards can be broadly categorised into two types: \textit{scattering billiards} (such as Sinai billiards\cite{YakovGSinai_1970} and Periodic Lorentz Gas\cite{Moran1987}) and \textit{defocusing billiards} (such as Bunimovich billiards\cite{Bunimovich1974}, Limaçon Billiard\cite{RevModPhys.85.869}, Polygonal Billiards with Rounded Corners\cite{harris2007polygonal}). The reflection from a fixed obstacle (scatterer having a convex curvature) placed within the billiard boundary always results in angle widening of the initially proximate trajectories (the green and purple trajectories in \cref{fig1}). This is fundamental to the chaotic scattering observed in billiards. The defocusing mechanism is an alternative to scattering, where a particle hits a convex boundary\cite{Wojtkowski1986}. This behaviour stands in contrast to the focusing mechanism, where reflections from concave boundaries result in regular, integrable dynamics. Walls with zero curvature are characterised by their flat nature, resulting in reflections that are non-chaotic (e.g., triangular, square, hexagonal or other polygonal billiards).Therefore, the nature of reflections in billiards strongly depends on wall geometry\cite{chernov2006chaotic, Wojtkowski1986}. The more irregular the boundary, the more sensitive the dynamical behaviour is to ICs.

From particle's point of view, the notion of curvature is from inside the billiard. Mathematically, we assume the sign of curvature ($\kappa$) as follows:
\begin{equation}\label{eq3}
	\kappa={\begin{cases}
			0 & \text{If}\; \dd{\varOmega}~ \text{is flat} ~(\dd{\varOmega^{0}}), \\
			+ve &  \text{If}\; \dd{\varOmega}~ \text{is convex}  ~(\dd{\varOmega^{+}}),\\
			-ve & \text{If}\; \dd{\varOmega}~ \text{is concave} ~(\dd{\varOmega^{-}}).
	\end{cases}}
\end{equation}
We are considering this notion through out our work. Any billiard boundary that combines any two of the aforementioned curvatures is referred to as a mixed-curvature billiard.

The first study on mixed-curvature billiard is the Bunimovich stadium billiard\cite{Bunimovich1974} having flat and concave segments and exhibits fully chaotic dynamics. Subsequent works have also explored similar combinations of boundary curvatures, such as the desymmetrised Cassini oval\cite{PhysRevE.57.5397} and polygonal billiards with rounded corners\cite{FLeyvraz_1996, PhysRevE.103.L030201}. While the Africa\cite{MVBerry_1987, PhysRevLett.116.023901} and the Limaçon\cite{RevModPhys.85.869} billiards consider concave wall with sharp convex turns, the Lemon billiard\cite{PhysRevE.59.303} consider sharp concave turns. While the stadium, Lemon, Limaçon, and polygonal billiards possess geometric symmetries, the other two billiards lack any such symmetry. In most of these systems (excluding the fully chaotic stadium billiard), the dynamics exhibits coexisting regular and chaotic regions. These studies indicate that the mixed dynamical behaviour is often associated with sharply curved boundary segments combined with predominantly smooth concave or convex regions. Recently, there have been a series of works\cite{PhysRevB.107.144308, PhysRevB.104.064310, PhysRevResearch.5.043028}, modelling the quantum chaos in graphene. In their models, the authors consider billiards having periodic or almost periodic concave and convex curvatures (small radius of curvatures). These have either $3-$ or $4-$fold rotational symmetries. They reported the presence of quantum chaos and showed the role of symmetry.

We propose two distinct classes of billiard systems\textemdash bean- and peanut-shaped\textemdash each characterised by their curvature profiles and symmetry properties with no neutral segments. In these proposed billiards, the concave/convex segments vary smoothly to convex/concave segments. The defining distinction between these two billiards is the difference of symmetry: the bean-shaped billiard has one mirror symmetry, while peanut has two. This difference allows us to examine how symmetry influences both classical phase-space structures and quantum spectral features.

Our investigation pursues two interconnected objectives. First, we aim to characterise the classical dynamics through billiard flow diagrams, Poincaré sections in Birkhoff coordinates, and an analysis of trajectory ensembles. Second, we aim to investigate how these classical chaotic behaviour translates into the quantum regime. Especially, how the presence of mirror symmetries enhances the eigenstate scarring; how complexity and information scrambling manifest and grow in these systems. To address these questions, we employ a comprehensive set of diagnostic tools. Traditional statistical indicators, such as the \textit{Nearest Neighbour Spacing Distributions} (NNSD), the \textit{Level Spacing Ratio} (LSR), and the spectral staircase function, are used to characterise the universal spectral signatures of chaos. Complementing these, we utilise two recently developed dynamical probes: \textit{Spectral Complexity} (SC) and the \textit{Out-of-Time-Order Correlator} (OTOC). 

The remainder of this paper is organised as follows. \cref{sec2} introduces billiard models with mixed-curvatures, utilising the Bean curve and Cassini curve, detailed in subsections \ref{sec2A} and \ref{sec2B}, respectively. Next, \cref{sec3} presents a classical analysis of billiard dynamics, covering the billiard flow, map, and Lyapunov exponent in subsections \ref{sec3A}, \ref{sec3B}, and \ref{sec3C} respectively. This section extensively examines the manifestation of classical chaos by analysing trajectories and phase space structures sensitive to ICs which highlight the underlying complex dynamics in these mixed-curvature billiards. Building on this foundation, \cref{sec4} provides a quantum mechanical perspective, demonstrating evidence of eigenfunction scarring within both chaotic billiard models in subsection \ref{sec4A}. We quantitatively analyse NNSD, LSR, and the spectral staircase function for four billiard configurations in subsections \ref{sec4B}, \ref{sec4B1}, and \ref{sec4B2}, respectively. Subsections {\ref{sec4C}} and {\ref{sec4D}} further explore the SC and OTOC of these billiards, where we find strong alignment with the classical results. Finally, \cref{sec5} offers a comprehensive summary of our findings. 

\section{\label{sec2} Models}
The quartic curve\cite{gibson1998elementary} serves as the foundational boundary function for our billiard boundary models. In algebraic geometry, a quartic plane curve is a fourth-degree curve on a plane. It can be defined by a bivariate quartic equation: 
\begin{widetext}
	\begin{equation}\label{eq4}
		A x^{4} + By^{4} + Cx^{3}y + Dx^{2}y^{2} + Exy^{3} + Fx^{3} + Gy^{3} + Hx^{2}y + Ixy^{2} + Jx^{2} + Ky^{2} + Lxy + Mx + Ny + P=0
	\end{equation}
\end{widetext}

In this investigation, two sets of coefficients in the aforementioned equation are analysed for two families of curves, namely bean curves and Cassini ovals, as outlined below. Symmetrical properties and relatively simple mathematical definitions make them highly significant for both theoretical studies and practical applications. 

\subsection{\label{sec2A}Bean Curves}
\begin{figure}[hbt!]
	\centering
		\includegraphics[width=\linewidth]{"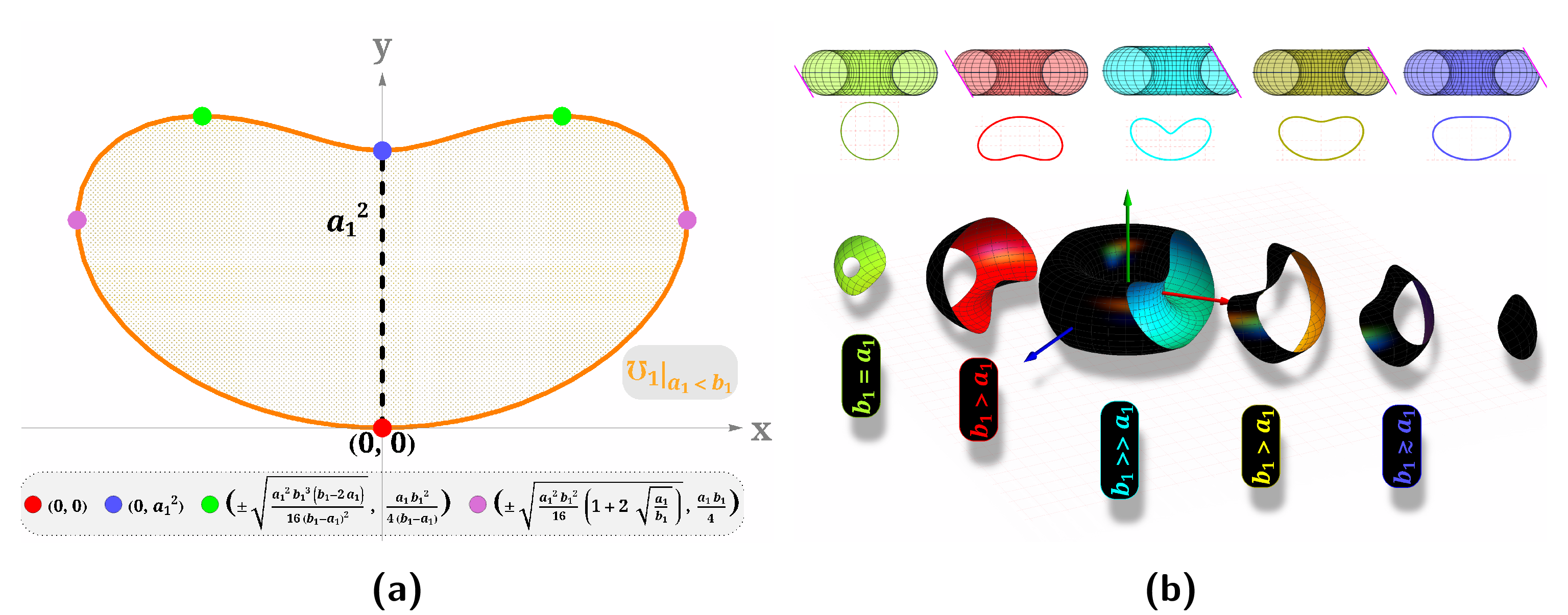"}
	\caption{(a) Bean curve for $ a_{1} < b_{1} $. The red and blue points are intercept points on the closed curve, whereas green and orchid points are extrema points. (b) Bean curves as planar sections of a torus. The top row features $ 2D $ slices of a torus, which take on different shapes from circular to bean-like curves as the parameters $ a_{1} $ and $ b_{1} $ are adjusted. In the bottom row, the torus is shown in $ 3D $, with each toroidal shape representing a different combination of $ a_{1} $ and $ b_{1} $.}
	\label{fig2}
\end{figure}

The Bean curves, discovered by Cundy and Rowllet\cite{cundy1961mathematical}, represent a specific quartic plane curve known for their distinctive bean-like appearance, as illustrated in \cref{fig2}(a). With genus\footnote{The genus of a surface refers to the number of ``holes'' it contains. For example, a torus has genus $1$, while a sphere has genus $0$.} zero, it has a singularity at the origin and a triple point.

The curve defining this billiard region is given by,
\begin{equation}\label{eq5}
	\mho_{1}(x, y, a_{1}, b_{1}) \coloneq (x^2 + y^2)^2 - a_{1} y(b_{1} x^2 + a_{1} y^2)=0
\end{equation}
Here, with two sets of parameter values we have two different shapes of the billiard as follows
		\begin{equation}\label{eq6}
			\mho_{1}(x, y, a_{1}, b_{1})=  
			\begin{cases}
				\text{Circle,} & \text{for}~ a_{1}=2~ \&~  b_{1}=2,\\
				\text{Bean,} & \text{for}~ a_{1}=2~ \&~ b_{1}=6 \\
			\end{cases}
		\end{equation}
Here, the circular billiard, as we know, has an entirely concave boundary, while the bean-shaped billiard has both concave and convex boundaries.

In contrast to the standard unit bean curve (also known as the egg curve)\cite{vaze2005continuous, wolframBeanCurve, weisstein2002crc}, which is located in the first and fourth quadrants, the implicit function denoted as $ \mho_{1} (x, y) $ in \cref{eq6}, for the bean-shaped billiard defines the boundary in the first and second quadrants. 

The \cref{fig2}(b) highlights the mathematical evolution of the bean curve and its geometric relationship to a torus. A torus intersected by a plane tangent to its axis reveals various bean curve forms. These toric-sections are influenced by the value of $ \dfrac{b_{1}}{a_{1}} $. The geometrical characteristics of bean curves are presented in \cref{table1}. We will only consider two forms of these curves, shown in \cref{eq6}, as a full exploration is beyond the present scope.

	\begin{table*}[hbt!]
		\centering
\caption{\label{table1} Geometrical properties of Bean and Cassini curves}
		\begin{tabular}{ rcc } 
			\toprule[0.07cm]
			\midrule
			\texttt{Properties} & \texttt{Bean curves} & \texttt{Cassini ovals}\\
\midrule
			\texttt{\textit{Intercepts:}}  & $ (0,~0 ) $, $ (0,~ a_{1}^2)$ & $ (\pm\sqrt{a_{2}^2\pm b_{2}^2},~0 ) $, $ (0,~\pm\sqrt{b_{2}^2-a_{2}^2} ) $\\
\rule{0pt}{8ex}
		\multirow[c]{2}{*}{\texttt{\textit{Extrema:}}}  & $ (0,~ a_{1}^2) $, $\bigg(\pm\sqrt{\dfrac{a_{1}^{2}b_{1}^{3}(b_{1}-2a_{1})}{16 (b_{1}-a_{1})^{2}}},~ \dfrac{a_{1} b_{1}^{2}}{4 (b_{1}-a_{1})}\bigg)^{*}$, & $\big(\pm\sqrt{a_{2}^{2}\pm b_{2}^{2}}, 0 \big) $, $ \big(0,~ \pm\sqrt{b_{2}^{2}-a_{2}^2}\big)^{*}$, \\&
		$\bigg(\pm\sqrt{\dfrac{a_{1}^{2} b_{1}^{2}}{16}(1+2\sqrt{\dfrac{a_{1}}{b_{1}}})},~ \dfrac{a_{1} b_{1}}{4}\bigg)$ & $\bigg(\dfrac{\pm\sqrt{4 a_{2}^{4}-b_{2}^{4}}}{2 a_{2}},~ \pm\dfrac{b_{2}^2}{2 a_{2}}\bigg) $ \\
\rule{0pt}{8ex}
		\texttt{\textit{Symmetries:}}  &  $ x=0 $; $ (0,~0) $ & $ x=0 $; $ y=0 $; $ (0,0) $ \\
\rule{0pt}{8ex}
		\multirow[c]{3}{*}{\texttt{\textit{Loops:}}} & a single connected loop for $ (\pm a_{1}, ~ \pm b_{1}) $ & a single loop if $ a_{2}<b_{2} $ \\& three connected loops  for $ (\pm a_{1}, ~ \mp b_{1}) $ & the lemniscate of Bernoulli, if $ a_{2} = b_{2} $ \\& \textemdash & two disconnected loops, if $ a_{2} > b_{2}$\\
\rule{0pt}{8ex}
			\texttt{\textit{Nodes:}}  & $ (0,~\dfrac{a_{1}^{2}}{2}) $ if $ a_{1}=b_{1} $ & $ (0,0) $ if $ a_{2}=b_{2} $  \\
			\bottomrule[0.07cm]	
		\end{tabular}
	\end{table*}

\subsection{\label{sec2B}Cassini Ovals}
The Cassini ovals (also known as Cassini ellipses or Cassinian curves or ovals of Cassini) are quartic curves characterised by the locus of points $ P $ such that the product of the distances from $ P $ to two fixed points, $ F_{1} $ and $ F_{2} $ (separated by a distance of $ 2a_{2} $), is $ b_{2}^{2} $. Mathematically, this is expressed as:

\begin{equation}\label{eq7}
	\abs{PF_{1}}\cdotp	\abs{PF_{2}}=b_{2}^{2}
\end{equation}
where $ b_{2} $ is a constant. The implicit function\cite{vaze2005continuous, wolframCassiniOvals}, $ \mho_{2} (x, y) $, defining a Cassini oval in Cartesian coordinates, with foci at $ (\pm a_{2}, ~0) $, is:
\begin{equation}\label{eq8}
	\mho_{2}(x, y, a_{2}, b_{2}) \coloneq (x^{2}+y^{2})^{2}-2 a_{2}^{2} (x^{2}-y^{2})+a_{2}^{4}-b_{2}^{4}=0
\end{equation}

Here also, we are considering two sets of parameter values that leads two different shapes of the billiard as follows:
\begin{equation}\label{eq9}
	\mho_{2}(x, y, a_{2}, b_{2})=
	\begin{cases}
		\text{Ellipse,} & \text{for}~ a_{2}=1~ \&~ b_{2}=10,\\
		\text{Peanut,} & \text{for}~ a_{2}=1~  \&~ b_{2}= \sqrt{1.375} \\
	\end{cases}
\end{equation}
Here, the elliptical billiard, with eccentricity $ 0.14072$, has an entirely concave boundary, while the peanut-shaped billiard has both concave and convex boundaries.

\begin{figure}[hbt!]
	\centering
	\includegraphics[width=\linewidth]{"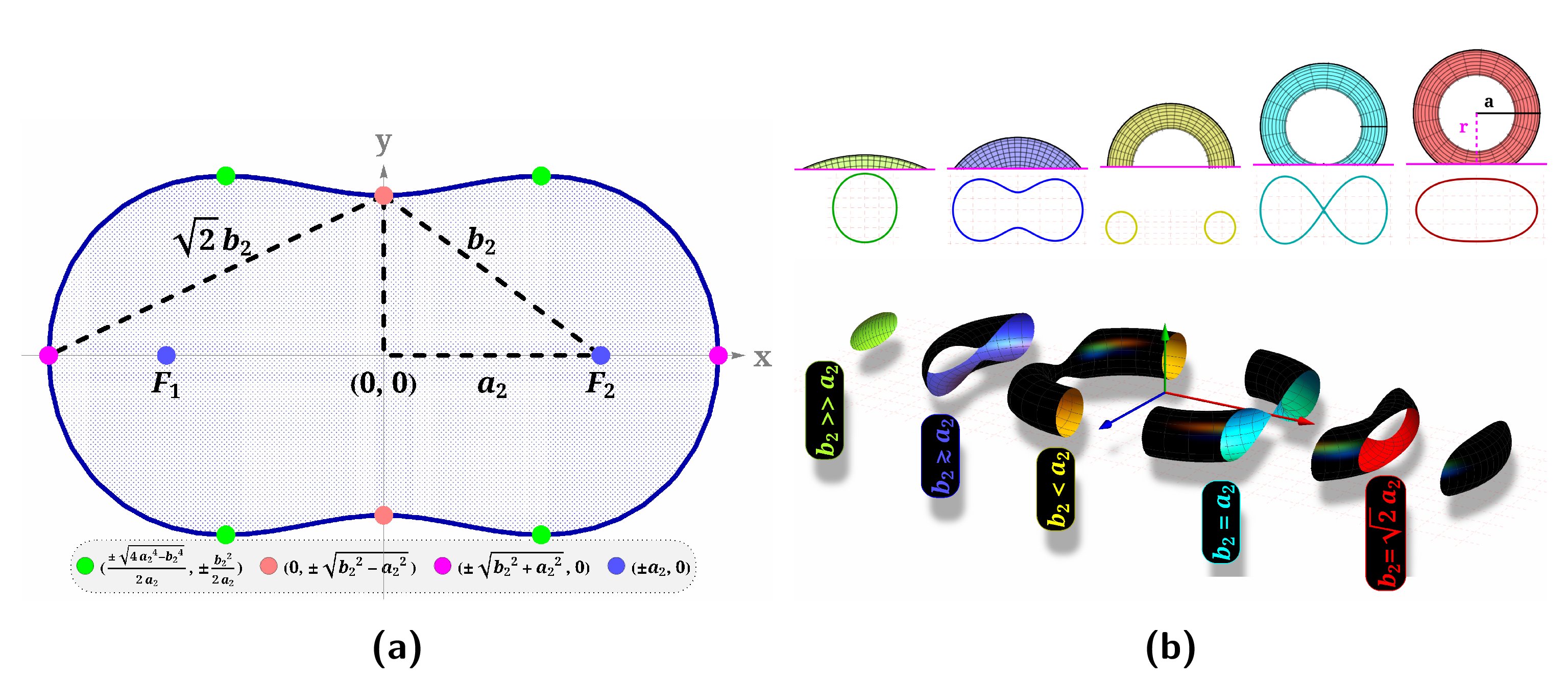"}
	\caption{\label{fig3} (a) Cassini oval with two foci ($ F_1 $ \& $ F_2 $) at $ (a_{2},0) $ and $ (-a_{2},0 $), respectively. The pink, magenta and green points are extrema points on the closed curve. (b) A family of Cassini Ovals as planar sections of a torus. The top row features $ 2D $ slices of a torus, whereas in the bottom row, the torus is shown in $ 3D $, with each toroidal shape representing a different combination of $ a_{2} $ and $ b_{2} $.}
\end{figure}

The geometrical properties of Cassini ovals are presented in \cref{table1}. The shape of these curves depends on $ \frac{b_{2}}{a_{2}}$. Different shapes of Cassini's ovals can also be found from the cross-sections of a circular torus cut by a plane parallel to its axis, which are now known as the ``spiric sections of Perseus''\cite{brieskorn2012plane, wells1991penguin, needham2023visual}. The details of these curves are beyond the scope of this work. Therefore, we shall limit ourselves to a rough picture of their forms\cite{moroni2017toricsectionssimpleintroduction} as shown in \cref{fig3}(b).

Having defined the geometry of the bean and peanut billiards, we now turn to their classical dynamics.

\section{\label{sec3} Classical analysis}

Classical chaos refers to a phenomenon in which deterministic systems\textemdash governed by precise laws\textemdash are extremely sensitive to tiny errors in ICs, causing arbitrarily close trajectories to separate rapidly\cite{strogatz2016nonlinear}. A notable example of such chaotic systems can be found in physical and mathematical billiards, which are classified as Hamiltonian systems. These systems have a natural invariant measure, preserving the ``volume'' in phase space.

To analyse the late-time dynamics, we utilise two complementary approaches: the continuous billiard flow and the discrete billiard map\footnote{A billiard map is a discrete-time map that captures the state of the system immediately after each reflection off the boundary.} (or Poincaré map). The latter serves as a more efficient tool for visualising phase-space structures. Here, rather than constantly monitoring the particle's position and velocity on the frictionless surface, this map reduces the phase space to a simpler form, focusing only on the boundary of a billiard and the angles of reflection. Despite this reduction, the essential dynamics of the system remain intact, allowing for a more straightforward yet still profound analysis of the chaotic nature of such Hamiltonian systems. 

The short-time dynamics, examined using an ensemble of independent single-particle trajectories, also demonstrate the dominant dynamical behaviour of the bounding geometry. In particular, the formation, deformation, or disappearance of caustics\footnote{Caustic is a curve to which reflected trajectories are tangent (for once or for always), creating concentrated geometrical patterns. In the case of a single IC (i.e. motion of a single particle), the reflected trajectories remain always tangent to the same curve after each reflection. However, when considering an ensemble of ICs (i.e. simultaneous evolution of many independent trajectories), these trajectories become tangent to different curves after each reflection.}\cite{Bruce1984, Hungerbühler27052020, BERRY1980257, gutzwiller2014chaos} over successive reflections provides a clear signature of how the boundary shape governs the underlying dynamics.

We are now ready to define the concept of a dynamical billiard. Consider a continuous curve $ \mathbf{x}(t)$, $t \in [0, \infty) $, in $ \varOmega $ with the following properties
\begin{enumerate}[
	align=left, leftmargin=\widthof{[Step-I]}+\labelsep]
	\item [\textit{Boundary conditions}:] The initial condition $ \mathbf{x}(0) \in \varOmega $. This means the particle remains within the domain throughout its motion.
	\item [\textit{Piecewise linear}:] $\forall~ t > 0 $, 
	$ \mathbf{x} $ consists of linear segments where each segment has its endpoints on $ \mho $. This means that between collisions with the boundary of $\varOmega$, the trajectory is linear.
	\item [\textit{Specular reflection}:] $\forall~ t > 0 $, the law of reflection holds good, i.e. $ \theta_{i}=\theta_{r} $.
	\item [\textit{Continuity and smoothness}:] $ x(t) $ is continuous and smooth within $ \partial{\varOmega}~ \forall~  t\in[0,\infty)$, except at the points of reflection where, $ x(t)~\in~\partial{\varOmega} $.
\end{enumerate}
Any trajectory $\mathbf{x}(t)$ that satisfies the aforementioned properties is termed as a ``billiard trajectory''.

For a particle moving within a billiard, the ICs comprise its starting position, speed, and direction. We can understand how the system evolves using either a billiard flow and/or a billiard map. The flow describes the particle's continuous movement. In contrast, the map offers a discrete-time perspective, examining the particle's state just after each boundary collision. The billiard flow is not smooth or differentiable; collisions with the boundary $\mho$ cause sudden changes in the particle's direction. These abrupt shifts highlight the system's inherent chaotic behaviour, where the tiniest variations in ICs result in dramatically different trajectories.

\subsection{\label{sec3A}Billiard flow}
The motion of a particle within the billiard is governed by a billiard flow, which characterises the sequence of reflections off the boundary. This flow provides a complete description of the particle's trajectory by specifying how each reflection leads to the next. The concept of the billiard flow is inspired by an optical analogy, where the phase space is viewed as a space of possible states, and the reflections represent transitions between these states. Through this optical perspective, the billiard flow captures the essential dynamics of the system, offering a comprehensive understanding of the particle's motion as it interacts with the boundary of the billiard.

Let $ \mathbf{x} \in \varOmega$ denote the position of the moving particle and $\mathbf{v}\in \mathbb{R}^2$ its velocity vector. Of course, $\mathbf{x} = \boldmath{x}(t)$ and $ \mathbf{v} = \mathbf{v}(t)$ are functions of time $t \in \mathbb{R}$. The particle moves with constant velocity between collisions and in between collisions, its motion follows straight lines, so
\begin{equation}\label{eq10}
\mathbf{x(t)}=\mathbf{x}_{0}+\boldmath{v}(t-t_{0})
\end{equation}
where $\mathbf{x}_{0}$ is the initial position, and $t_{0}$ is the initial time. Let $n$ be the unit normal vector at the point of collision with the billiard wall. If $\mathbf{v}_{in}$ refers to the pre-collisional velocity, then the post-collisional $\mathbf{v}_{out}$ is given by,
\begin{equation}\label{eq11}
\mathbf{v}_{out}= \mathbf{v}_{in} -2(\mathbf{v}_{in} \cdot \mathbf{n})\cdot \mathbf{n}
\end{equation}
This reflection law ensures the particle's speed remains constant, but its direction changes depending on the geometry of the boundary.


\subsubsection{Single particle dynamics}

\begin{figure}[hbt!]
	\centering
	\includegraphics[width=\linewidth]{"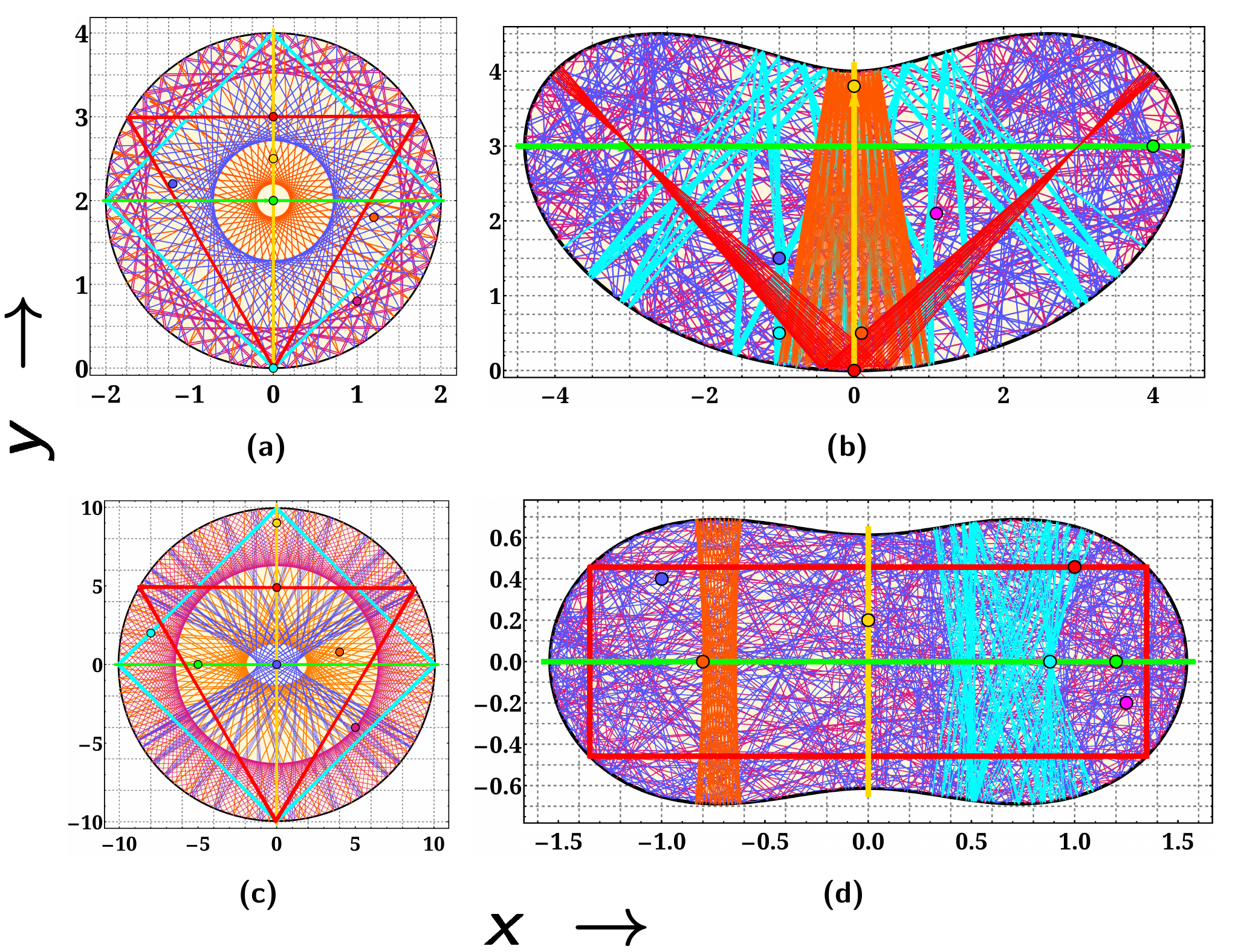"}
	\caption{\label{fig4} Billiard flow diagrams representing real space trajectories (periodic, quasi-periodic and chaotic) for (a) Circle, (b) Bean, (c) Ellipse, and (d) Peanut billiards for different ICs. Here, $\{{\color{green} \CIRCLE}, ~{\color{gold(web)(golden)}\CIRCLE}, ~{\color{red}\CIRCLE}, ~{\color{cyan} \CIRCLE}, ~{\color{orange}\CIRCLE}, ~{\color{blue}\CIRCLE}, ~{\color{magenta!70} \CIRCLE}\}$ represent seven initial positions. The reflection trajectories within circular and elliptical billiards are predictable, periodic, or quasi-periodic. Within the boundaries of a circular billiard, every concentric circle acts as a caustic. Conversely, within the elliptical billiard, all confocal ellipses and hyperbolas are caustics. Chaotic trajectories are the norm for bean- and peanut-shaped billiards, with only a few specific ICs resulting in periodic or quasi-periodic outcomes.}
\end{figure}

In classical billiards with either concave or flat boundary, a particle's trajectories are periodic\textemdash meaning it eventually retraces its path after a finite number of reflections. However, when the billiard has a mixed-curvature boundary (a combination of two or more different curvature segments), the nature of dynamics changes. Despite the system still being deterministic, the particle's path may never repeat. Each collision alters the particle's course, making its future trajectory highly sensitive to ICs, leading to chaos. Crucially, these chaotic billiards exhibit both hyperbolicity (nearby trajectories diverge exponentially) and ergodicity (a typical path eventually covers the entire space).

In circular and elliptical billiards, shown in {\cref{fig4}}(a \& c) respectively, the reflections for different ICs result in periodic, or quasi-periodic trajectories. Here, the term ``periodicity'' refers to the number of boundary reflections required for the trajectory to return to its IC, both in position and direction\cite{chernov2023chaotic, doi:10.1098/rsta.2017.0419}. The yellow and green trajectories in the circular and elliptical billiards illustrate period-two oscillations, characterised by a simple back-and-forth motion where the particle reverses its velocity vector upon each collision with the boundary. Other examples of periodic motion are shown in {\cref{fig4}}(a \& c), where the particle traverses the sides of some regular polygons, forming period-three triangle (red colour), and period-four square (cyan colour) etc. Except for these, depending on the choice of ICs, caustics are formed in these billiards indicating integrability of their dynamics. In circular billiard, concentric circular caustics are formed, which are depicted in orange, blue and magenta colours. Moving the ICs farther from the centre of circle increases the radius of the circular caustics. Unlike the circular billiard, there are two kinds of caustics in the elliptical billiard sharing the same foci $F_{1}$ and $F_{2}$. Here, outer trajectories (trajectories whose paths fall outside the focal points after each reflections) produce confocal elliptic caustics (magenta and orange), while inner trajectories (trajectories whose paths fall inside the focal points after each reflections) result in confocal hyperbolic caustics (blue). Producing these caustics from a single IC involves multiple reflections.

\cref{fig4}(b \& d) show both regular and chaotic trajectories in the bean- and peanut-shaped billiards, respectively. By selecting the appropriate ICs, we can identify subregions within these boundaries where trajectories are constrained to specific areas, resulting in periodic or quasi-periodic trajectories. In the bean-shaped billiard, the yellow and green trajectories correspond to period-two oscillations. In contrast, the yellow, green and red trajectories in the peanut-shaped billiard, though appear periodic, exhibit slow diverging motion. The red, cyan, and orange trajectories in the bean-shaped billiard, along with the orange and cyan ones in the peanut-shaped billiard, represent quasi-periodic motion trapped in those subregions within the mixed phase space. Any other ICs (shown in blue and magenta) outside these subregions cause trajectories to become non-periodic, filling the space erratically, reflecting chaotic behaviour in both the mixed-curvature billiards. These results stem from the hyperbolic and ergodic features of chaotic billiards. Unlike regular billiards, single particle dynamics does not produce any caustics here.

\subsubsection{Caustic formation by ensembles of trajectories}

\begin{figure}[hbt!]
	\centering
	\includegraphics[width=\linewidth]{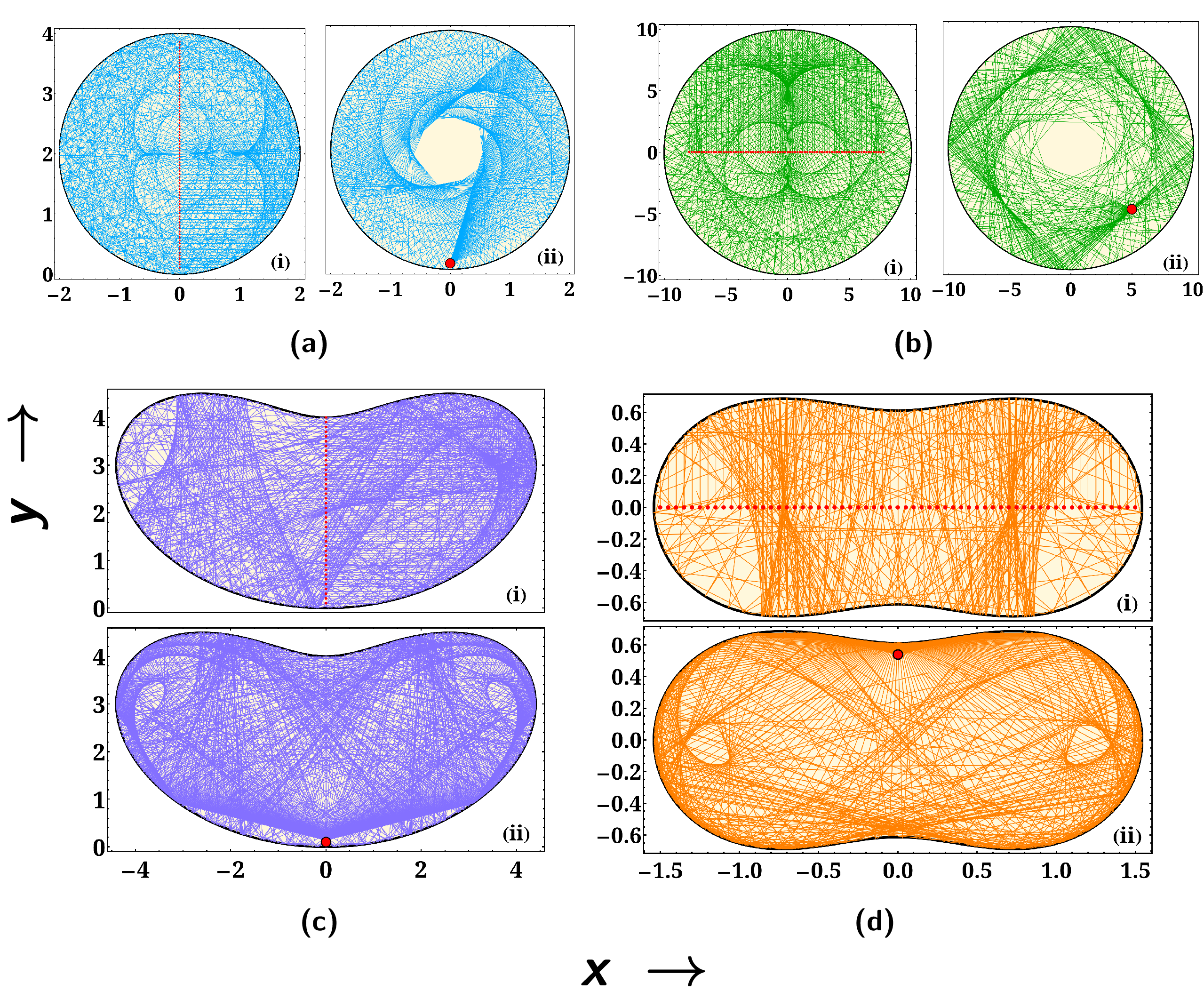}
	\caption{Caustic patterns produced by an ensemble of independent trajectories in (a) Circle, (b) Ellipse, (c) Bean and (d) Peanut billiards. In (a-d), sub-figures labelled (i) correspond to the first sets of ICs, where only the positions differ and sub-figures labelled (ii) represent the second set of ICs characterised by variations in the direction of momenta. Here, the {`${\color{red}\CIRCLE}$'} represent the ICs. In circular and elliptical billiards, regular geometry (constant curvature) enforces a regular focusing effect, causing the trajectories to trace out recurrent caustics. Whereas, the mixed-curvature breaks the symmetry and continuity of caustic formation, resulting in non-repetitive, irregular trajectories and the destruction of stable caustics. This geometric complexity gives rise to chaotic behaviour.}
	\label{fig5}
\end{figure}

To statistically probe the local stability of the boundary, we now consider an ensemble of independent particles (or, equivalently, a bundle of independent trajectories) launched simultaneously. The equal-time evolution of these trajectories represents numerous independent realisations of the same one-particle Hamiltonian. Therefore, they more accurately reflect an ensemble of ICs, rather than a true many-body system. Here, we are considering two sets of such ICs: 
\begin{enumerate}[label={(\roman*)}]
	\item Different positions with identical momentum,
	\item Same position with different momentum (same magnitude but different directions).
\end{enumerate}
Since the particles do not interact, each trajectory evolves independently under the same Hamiltonian. The advantage of this approach is that it generates local caustics after every reflection, in contrast to the earlier case, where several reflections were required for their formation. To prevent overcrowding, we have limited the analysis to a maximum of five reflections per billiard.

\cref{fig5} shows different caustics patterns in the billiards for the above-mentioned two sets of ICs. For circular and elliptical billiards (\cref{fig5}(a-i \& b-i)), because of their smooth concave boundary, the first set of ICs produce cusp-caustics. While the second set of ICs produce almost indistinguishable patterns (folds and cusps) in both these billiards as shown in \cref{fig5}(a-ii \& b-ii). These well organised and repetitive patterns correspond to stable structures or constant curvature of the boundary reflecting integrability of the system\cite{MVBerry_1979}. When these boundaries include convex segments, the previously seen repetitive patterns for both sets of ICs dissolve into more complex and non-repetitive patterns. \cref{fig5}(c \& d) show various such patterns for bean- and peanut-shaped billiards, respectively. In both these billiards, concave boundary produces converging trajectories leading to cusp like caustics, while the particles bouncing from both concave and convex boundaries produce smoothly varying, broadened and less intense patterns. The presence of non-repetitive caustics in mixed-curvature billiard systems offers compelling evidence for the breakdown of integrability, indicating the emergence of chaotic dynamics.




\subsection{\label{sec3B} Billiard map}
By using the Poincaré \textit{surface of section} (SOS) method, classical billiard dynamics can be simplified to a two-dimensional discrete mapping. This method utilises the boundary of the billiard table as the SOS to discretise the dynamics. A billiard flow has a natural Poincaré section defined by Birkhoff coordinates\cite{ChaosBook}. These coordinates are the arc-length position of the $ n^{th} $ bounce along the billiard boundary ($ \partial{\varOmega} $), denoted by $ \xi $, and the tangential component of momentum at the boundary, denoted by $ p_{n} = \abs{p} \sin(\phi_{n}) $. Here, $ \phi_{n} $ represents the angle between the outgoing trajectory and the normal to the boundary \cite{LeGhAl13ch5, ChaosBook, sym17020202, Bunimovich02102021, Leonel2021}. Both the arc length $\xi$ and the tangential momentum $p$ are measured counter-clockwise relative to the outward normal. Here, each point in the Poincaré section represents a collision event characterised by the location of the particle on the boundary and the angle of the reflection. Over time, as the particle bounces around, a collection of points forms the collision space with coordinates $\phi$ and $\xi$, which we call the Poincaré map or the collision map $\mathcal{P}$.

Liouville's theorem ensures that volume is preserved in full phase space. Given the straight-line motion within the billiard, it is practical to use the boundary to define a Poincaré section. Mathematically,
\begin{equation}\label{eq12}
\mathcal{P} \coloneq {(\xi,p)~|~\xi\in[0,\abs{\partial{\varOmega}}],~p\in[-\abs{p},\abs{p}]}
\end{equation}
The dynamics from the $ n^{th} $ collision to the $ (n + 1)^{th} $ collision is given by
\begin{equation}\label{eq13}
\mathcal{P} : (\xi_{n}, sin(\phi_{n})) \mapsto (\xi_{n+1}, sin(\phi_{n+1}))
\end{equation}
where, $ \xi \in \big[0,~\abs{\partial{\varOmega}}\big] $ and $ \phi \in[0,~\pi] $. Note that, The collision map $\mathcal{P}$ also admits an involution, i.e. $(\xi, p) \mapsto (\xi, -p)$.

Regular orbits on a Poincaré section appear orderly and confined. They appear as points or closed curves, indicating periodic or quasi-periodic motion. On the other hand, chaotic trajectories scatter irregularly and densely fill the available areas in a disorderly manner. Chaotic regions often surround the subregions dominated by quasi-periodic orbits. This coexistence creates a complex structure in phase space, where stable, regular regions are interwoven with chaotic zones\cite{Datseris2022}.

\begin{figure*}[hbt!]
\centering
\includegraphics[width=\linewidth]{"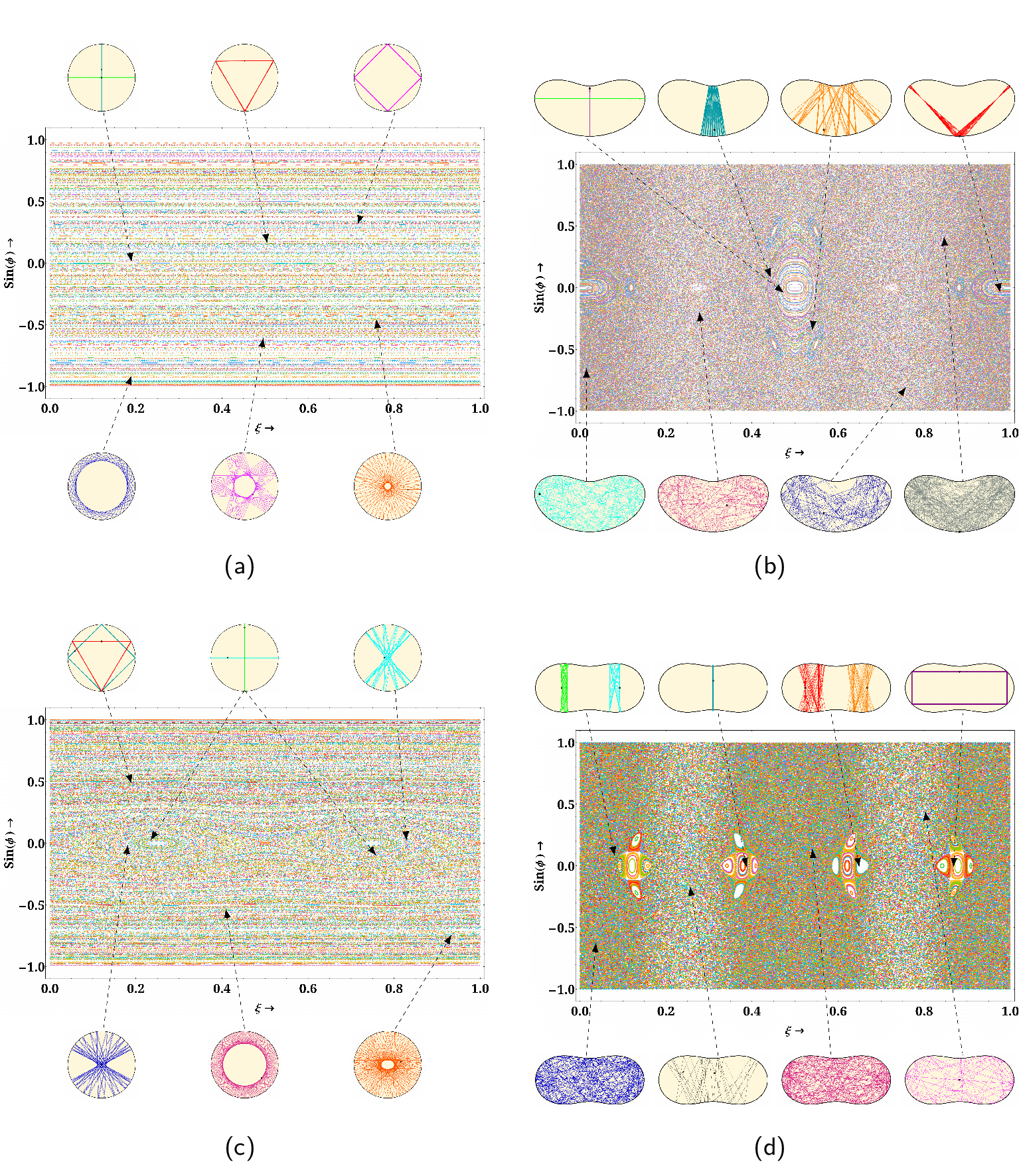"}
\caption{\label{fig6}Poincaré section of (a) Circle, (b) Bean, (c) Elliptical and (d) Peanut billiards, together with representative trajectories in real space. The smooth, well-confined invariant curves in the $(\xi, ~\sin(\phi))$ plane observed for the circular and elliptical billiards confirm their integrable nature. In contrast, the widespread scattering of points found in bean- and peanut-shaped billiards, along with pronounced variations in their density, clearly demonstrates the presence of dominant chaotic dynamics. Compared to the bean-shaped billiard, the peanut's greater geometric symmetry organises the stability islands into regular patterns, creating structure in the chaotic sea. The accompanying trajectories, displayed in position space, qualitatively indicate the location of the Birkhoff coordinates on the Poincaré section.}
\end{figure*}

Billiard Map for our chosen billiard boundaries are shown in \cref{fig6} along with a qualitative demonstration of their corresponding billiard flow diagrams. In \cref{fig6}(a), for the circular billiard, the data points create perfectly straight horizontal lines each at constant value of $\sin(\phi)$. Here, each line is densely populated with points, reflecting the conservation of angular momentum, as $\sin(\phi)$ is directly proportional to this conserved quantity. As a result, each horizontal line represents a unique invariant torus. Trajectories that are confined to these tori display either periodic or quasi-periodic motion. This behaviour is a direct consequence of the circle's constant and concave curvature, which produces a uniform focusing effect at the boundary. As a result, the dynamics supports a concentric family of circular caustics, each associated with a fixed value of angular momentum. In the Poincaré section, this regularity manifests as a set of perfectly ordered points, with no distortion or mixing between neighbouring tori. The absence of any fragmentation or scattering of points therefore confirms that the circular billiard is fully integrable, exhibiting high symmetry and no signatures of chaotic dynamics.

In contrast, the elliptical billiard (\cref{fig6}(c)), although integrable, exhibits a richer phase-space structure. Governed by the confocal geometry of the boundary, its dynamics give rise to smooth, nested invariant curves separated by a characteristic hyperbolic separatrix. Here, the values of $\sin(\phi)$ vary, but do so in a systematic, predictable, and repeating manner. A prominent feature of the Poincaré section is the appearance of an ``$\infty-$shaped'' closed curve on $\mathcal{P}$, separating two distinct regions of closed and open curves \cite{chernov2006chaotic}. The open curves, also known as ``horizontal waves'', which lie beyond the $\infty-$shape, correspond to all the trajectories tangent to the elliptic caustics. Meanwhile, the trajectories tangent to the hyperbolic caustics form closed loops inside the $\infty$-shape. These characteristics highlight that integrability in the elliptical billiard is associated with a complex and well-structured phase-space topology.

Introducing mixed curvature along the boundary qualitatively alters the overall outcome. In the bean-shaped billiard, the prominent feature in \cref{fig6}(b) is a broad region of irregularly scattered points, commonly referred to as the chaotic sea. This region indicates ergodic motion, with trajectories exploring much of the energy shell. Embedded within this sea are distinct, regular islands (smooth, closed curves). These islands correspond to surviving \textit{Kolmogorov-Arnold-Moser} (KAM) tori and host quasi-periodic motion localised around stable periodic orbits. The coexistence of these regions results from the competing focusing (concave) and defocusing (convex) segments of the boundary. The alternating convex and concave boundary segments break the rotational symmetry present in the circular billiard and destroy the associated conservation laws. As a result, invariant tori are only partially preserved, giving way to extended chaotic regions. Therefore, the visually dominant chaotic sea in the Poincaré section provides clear evidence of predominantly chaotic classical dynamics in the bean-shaped billiard.

This shift towards predominantly chaotic dynamics is more pronounced in the peanut-shaped billiard. In \cref{fig6}(d), the Poincaré section similarly comprises a chaotic sea interspersed with regular islands; however, the balance between these components is pushed further in favour of chaos. The chaotic sea occupies a substantially larger portion of phase space, indicating that a greater fraction of ICs give rise to chaotic trajectories. In contrast, the regular islands persist only as smaller, less numerous structures, reflecting the diminished stability of periodic orbits in this geometry. This redistribution of phase-space volume underscores the stronger chaotic character of the peanut-shaped billiard.

The distribution of collision points in the Poincaré section provides direct insight into the underlying phase-space dynamics. Dense clusters of points identify regions of phase space where trajectories spend a disproportionately long time, indicating partial confinement. In contrast, sparsely populated areas correspond to strongly hyperbolic regions of phase space. Trajectories passing through these regions diverge exponentially fast. Therefore, the Poincaré section records very few collisions $(\xi, ~\sin(\phi))$ points, leading to a low sampling density. The spatial arrangement and morphology of these high- and low-density patches differ between the bean- and peanut-shaped billiards. These differences directly reflect variations in the boundary curvature profiles, which modify the balance between focusing and defocusing mechanisms and, consequently, the local stability properties of the classical dynamics.

In summary, the classical phase space of mixed-curvature billiards exhibits a hierarchical structure in which chaotic seas coexist with regular islands, with their relative sizes depending on the boundaries and symmetries. The convex boundaries cause nearby trajectories to diverge, fuelling chaotic dynamics. Concave segments, in contrast converge nearby trajectories and preserve certain stable caustics, allowing for periodic or quasi-periodic orbits. Trajectories near the boundary between regular and chaotic regions often exhibit intermittency, where a particle spends a long time near regular islands before escaping into chaos. These serve as partial barriers. In the domains dominated by concave curvature, the Poincaré section typically reveals smooth invariant tori, indicating quasi-periodic motion and near-integrable dynamics. In contrast, regions shaped by convex curvature generate scattered points in the Poincaré section, reflecting chaotic behaviour and strong sensitivity to ICs. Therefore, the above study suggests that the two mixed-curvature billiard systems are predominantly chaotic.

\subsection{\label{sec3C}Lyapunov exponent}

To quantify the manifestation of chaos, specifically extreme sensitivity to changes in ICs, Lyapunov exponents are utilised\cite{PhysRevE.52.2401, Wojtkowski1986}. In billiard systems, a particle moves in a straight line between boundary reflections, during which nearby trajectories separate only linearly. This emphasises that boundary collisions are crucial to the development of chaos in billiards, as exponential instability originates exclusively at these points. For this reason, it is more natural to parametrise the dynamics using the collision index $k$, representing the number of reflections, rather than continuous time.

\begin{figure}[hbt!]
 	\centering
 	\includegraphics[width=\linewidth]{"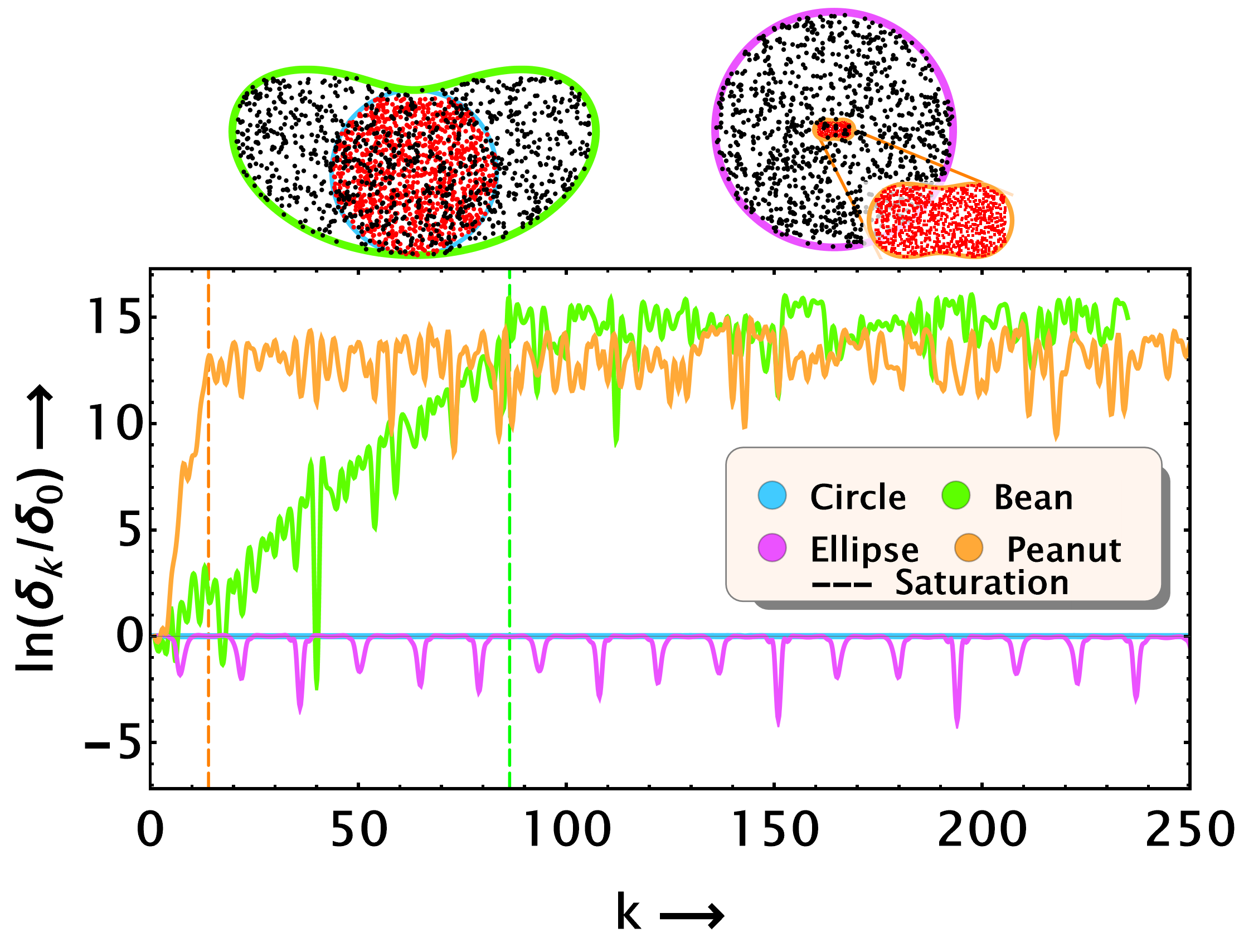"}
 	\caption{Separation growth between two initially close trajectories ($10^{-6}$) as a function of collision number $k$. For each billiard system, one representative pair was selected from a set of $1000$ trajectories, considering a maximum of $k_{\max} = 250$ collisions. The slope of the unsaturated region gives the Lyapunov exponent. The black and red markers in the upper inset show the randomly chosen ICs.}
 	\label{fig7}
 \end{figure}

Let $\delta_{k}$ denote the magnitude of the angular separation between two nearby trajectories after $k$ collisions, evaluated at the $(k+1)^{\mathrm{th}}$ impact. Accordingly, $\delta_{0}$ corresponds to their initial angular difference at the first collision. For a chaotic billiard, the angular separation grows exponentially with time, according to $\delta_{k}\sim\delta_{0}\mathrm{e}^{\lambda_{L}k} $. To quantify the divergence, one examines the growth of $\ln(\frac{\abs{\delta_{k}}}{\abs{\delta_{0}}})$ as a function of $k$ for representative trajectories (shown in \cref{fig7}). The Lyapunov exponent is then extracted as the slope of the linear, pre-saturation regime of this curve.

$d_{\textrm{avg}}$ characterises the typical free-flight length and thus the frequency of boundary interactions, which are the sole source of instability in billiard systems. A smaller $d_{\textrm{avg}}$ value implies more frequent collisions, leading to stronger cumulative divergence, while larger values correspond to slower instability build-up. The normalised quantity $\frac{d_{\textrm{avg}}}{\abs{\Omega}}$ removes system-size dependence and reflects how efficiently trajectories explore the available space. Lower $\frac{d_{\textrm{avg}}}{\abs{\Omega}}$ values indicate more collisions and stronger geometric influence. In contrast, the Lyapunov exponent is determined entirely by the geometry of the billiard table and is independent of the time or number of collisions ($k$) required for the trajectory separation to reach saturation (as shown in \cref{fig7}). For this reason, it is directly comparable to the geometric Lyapunov exponent introduced here.
 
To ensure statistical reliability, the procedure is repeated over $1000$ trajectories, each with different ICs, and the resulting Lyapunov exponents are averaged. While the magnitude of velocity does not influence Lyapunov exponent calculations in billiard systems, it is consistently fixed at unity for uniformity. In addition, the mean distance between successive collision points, $d_{\textrm{avg}}$, is computed for each trajectory and then averaged over the ensemble. The corresponding average Lyapunov exponents and mean collision distances for the billiards studied are listed in \cref{table2}.

	\begin{table}[hbt!]
		\centering
		\caption{\label{table2} Average Lyapunov exponents $(\lambda_{L})_{\textrm{avg}}$ and average distance between consecutive collisions $d_{\textrm{avg}}$.}
		\begin{tabular}{rccc} 
			\toprule[0.07cm]
			\midrule
			\texttt{Billiards} & $d_{\textrm{avg}}$ & $\frac{d_{\textrm{avg}}}{\abs{\Omega}}$\footnote{$\abs{\Omega}$ is the area of the billiard ref. \cref{table4}} & $(\lambda_{L})_{\textrm{avg}}$\\
			\midrule
			\texttt{\textit{Circle:}}  & $3.36$ & $0.267$ & $\sim0$\\
			\texttt{\textit{Bean:}}  & $4.25$ & $0.135$  & $0.079$\\
			\texttt{\textit{Ellipse:}}  & $1.70$ & $0.005$ & $\sim0$\\
			\texttt{\textit{Peanut:}}  & $1.48$ & $0.403$ & $0.029$\\
			\bottomrule[0.07cm]	
		\end{tabular}
	\end{table}

The \cref{table2} summarises the relationship between geometric properties of different billiard shapes and their dynamical behaviour. The integrable cases, namely the circle and ellipse, exhibit vanishing Lyapunov exponents, indicating regular, non-chaotic motion despite differing collision scales. In contrast, the bean- and peanut-shaped billiards display positive Lyapunov exponents, reflecting sensitivity to ICs and the onset of chaotic dynamics. The normalised $\frac{d_{\text{avg}}}{\abs{\Omega}}$ does not show a straightforward correlation with chaoticity, suggesting that boundary curvature and shape complexity play a more decisive role than simple geometric scaling.

\section{\label{sec4} Quantum Mechanical Analysis}

To develop a more comprehensive understanding of classical chaos, it is essential to investigate it through the lens of quantum mechanics \cite{Iitaka}. However, the classical approach is rendered infeasible by the Heisenberg uncertainty principle, which limits our ability to precisely measure both position and momentum simultaneously, forcing us to seek a different approach. Therefore, quantum chaos research focuses on examining the statistical characteristics of eigenfunctions and energy levels, rather than tracking the system's temporal dynamics.

The well-known linear Schrödinger equation:
\begin{equation}\label{eq14}
\mathcal{H}\psi(\mathbf{x})=\Bigg(\dfrac{-\hbar^{2}}{2m}~ \laplacian+V(\mathbf{x}) \Bigg)\psi(\mathbf{x})=\mathcal{E}\psi(\mathbf{x})
\end{equation}
Here, we adopt the units: $ \hbar = k_{\beta} = m = 1 $. Since potential inside the billiard is zero ($ V(\mathbf{x})=0~ \forall~ \mathbf{x}~ \in~ \varOmega $), this equation reduces to Helmholtz equation:
\begin{equation}\label{eq15}
\Bigg(\dfrac{-1}{2} ~\laplacian \Bigg)\psi(\mathbf{x})=k^{2}\psi(\mathbf{x}),\qquad \forall~ \mathbf{x} \in \varOmega
\end{equation}
with Dirichlet boundary conditions, i.e. $ \psi(\mathbf{x}) = 0$ $\forall$ $\mathbf{x} \in \partial{\varOmega}$. Here $ \laplacian $ denotes the Laplace operator, which reads in two dimensions $\laplacian = ( \pdv[2]{}{x_1}+ \pdv[2]{}{x_2}) $. Here, the eigenenergy $ \mathcal{E} = k^{2} $, where $ k $ is the wave number, and the interpretation of $\psi$ is that $ \int_{\mathcal{D}}\abs{\psi(\mathbf{x})}^{2}\dd[2]{\mathbf{x}} $ is the probability of finding the particle inside the domain $ \mathcal{D} \subset \varOmega $. In quantum billiards, we find the stationary solutions of the Schrödinger equation by determining the eigenvalues and eigenfunctions of the Helmholtz equation, which give us insights into the quantum equivalents of classical chaotic behaviour.

\begin{figure}[hbt!]
\centering
\subfigure[\label{fig8a}]{\includegraphics[width=\linewidth]{"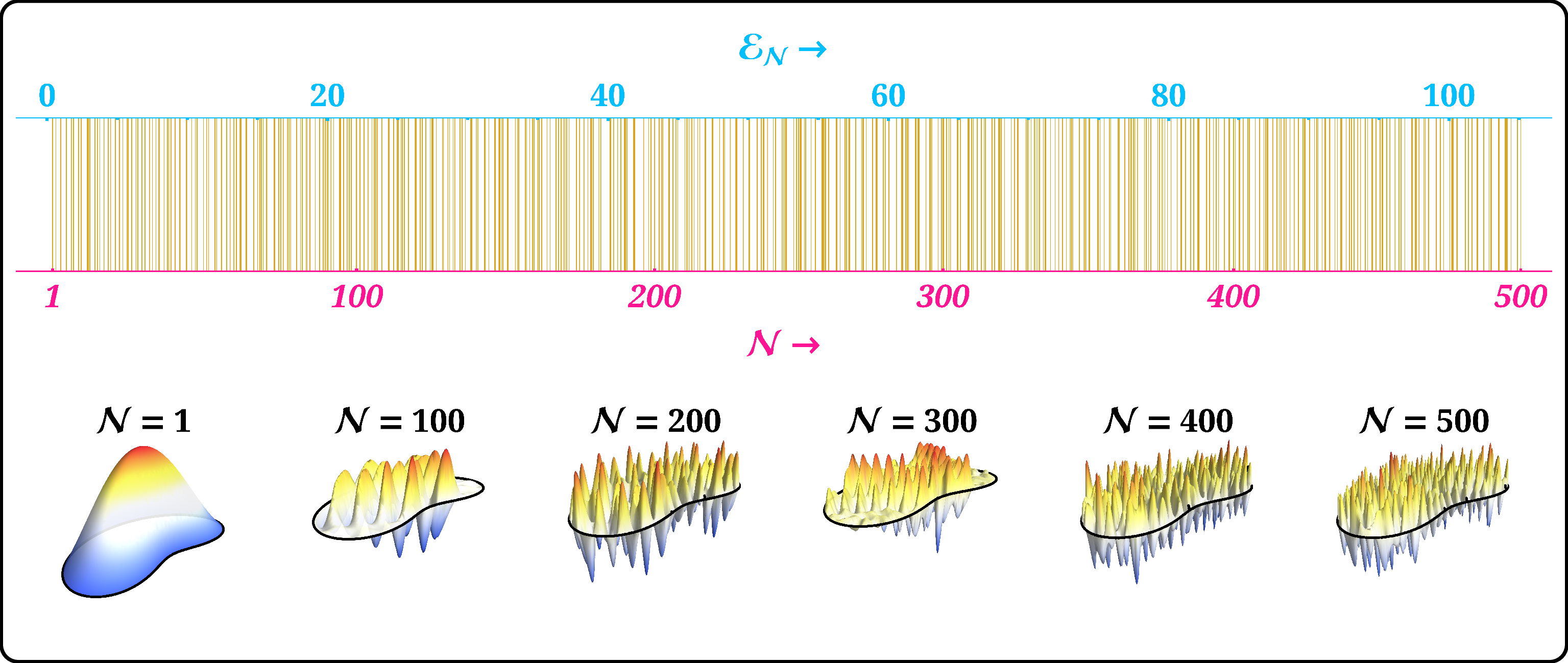"}}
\vfill
\subfigure[\label{fig8b}]{\includegraphics[width=\linewidth]{"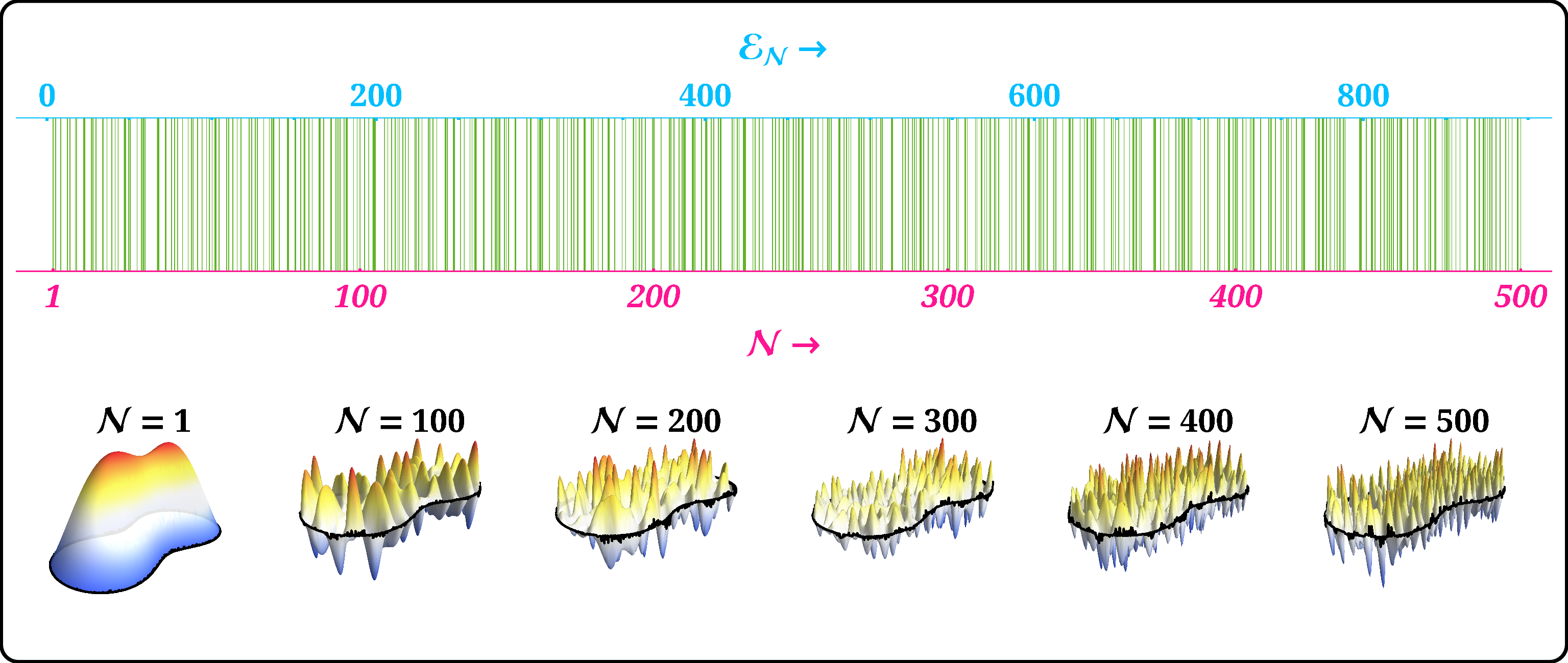"}}
\caption{\label{fig8}In (a) Bean and (b) Peanut billiards, the upper and lower rows represent first $500$ eigenvalues and a few selective eigenstates, respectively.}
\end{figure}

For certain simple domain, it is possible to solve {\cref{eq15}} analytically. However, numerical techniques such as \textit{Finite Difference Method} (FDM)\cite{MIKHAILOV1994375}, \textit{Finite Element Method} (FEM)\cite{Mikhailov1994}, \textit{Boundary Element Method} (BEM)\cite{beer2008boundary}, etc. are commonly employed for more complex domains. In this work, we have employed FEM in \texttt{Mathematica $14.0$} to compute around $ 4 \times 10^{4} $ eigenvalues for all the models under consideration. {\cref{fig8a}} displays the numerical outcomes for the Bean-shaped billiards, while {\cref{fig8b}} illustrates the results for the peanut-shaped billiards. The vertical lines (yellow and green lines) indicate energy eigenvalues $\mathcal{E}_{\mathcal{N}}$ corresponding to their quantum numbers $\mathcal{N}$, with $\mathcal{N}$ ranging from $1$ to $500$. The spacing between the eigenvalues appears small, but non-uniform. This irregularity in spacing provide insights into the nature of the system's behaviour, which will be further explored in the following sections. \cref{fig8a,fig8b} also shows the a few selected eigenstates corresponding to different $ \mathcal{N} $. The colour gradient likely represents the amplitude of the eigenfunction (red for high amplitude, blue for low amplitude), with the wavefunction oscillating between positive and negative values. The shapes of these eigenstates directly relate to the potential governing the system, with each eigenstate reflecting the spatial distribution of the probability density for a particle.

\subsection{\label{sec4A}Evidence of scarring}
In quantum mechanics, the wavefunction ($\psi_{\mathcal{N}}$) plays a crucial role in determining how particles behave. Therefore, while translating ergodicity to the quantum realm, it is logical to assume that the corresponding $\psi_{\mathcal{N}}$ of a classical ergodic system would spread uniformly over the energy shell. This is the essence of the ``quantum ergodicity theorem''\cite{shnirel1974ergodic, Shnirelman2023, zelditch2005quantumergodicitymixing, sunada1997quantum}, which holds for almost all eigenstates. However, exceptions arise in quantum chaotic systems: some eigenstates exhibit quantum scarring\cite{PhysRevLett.53.1515, Turner2018, PhysRevB.106.205150, PhysRevLett.132.020401}, where the wavefunction concentrates along unstable classical periodic orbits(``slow diverging'' or ``near marginally stable''). Even though quantum systems typically evolve toward thermal equilibrium, effectively erasing the memory of their ICs\cite{PhysRevE.50.888}, scarred states are more likely to stay close to their initial periodic orbits.

\begin{figure}[hbt!]
	\centering
	\includegraphics[width=\linewidth]{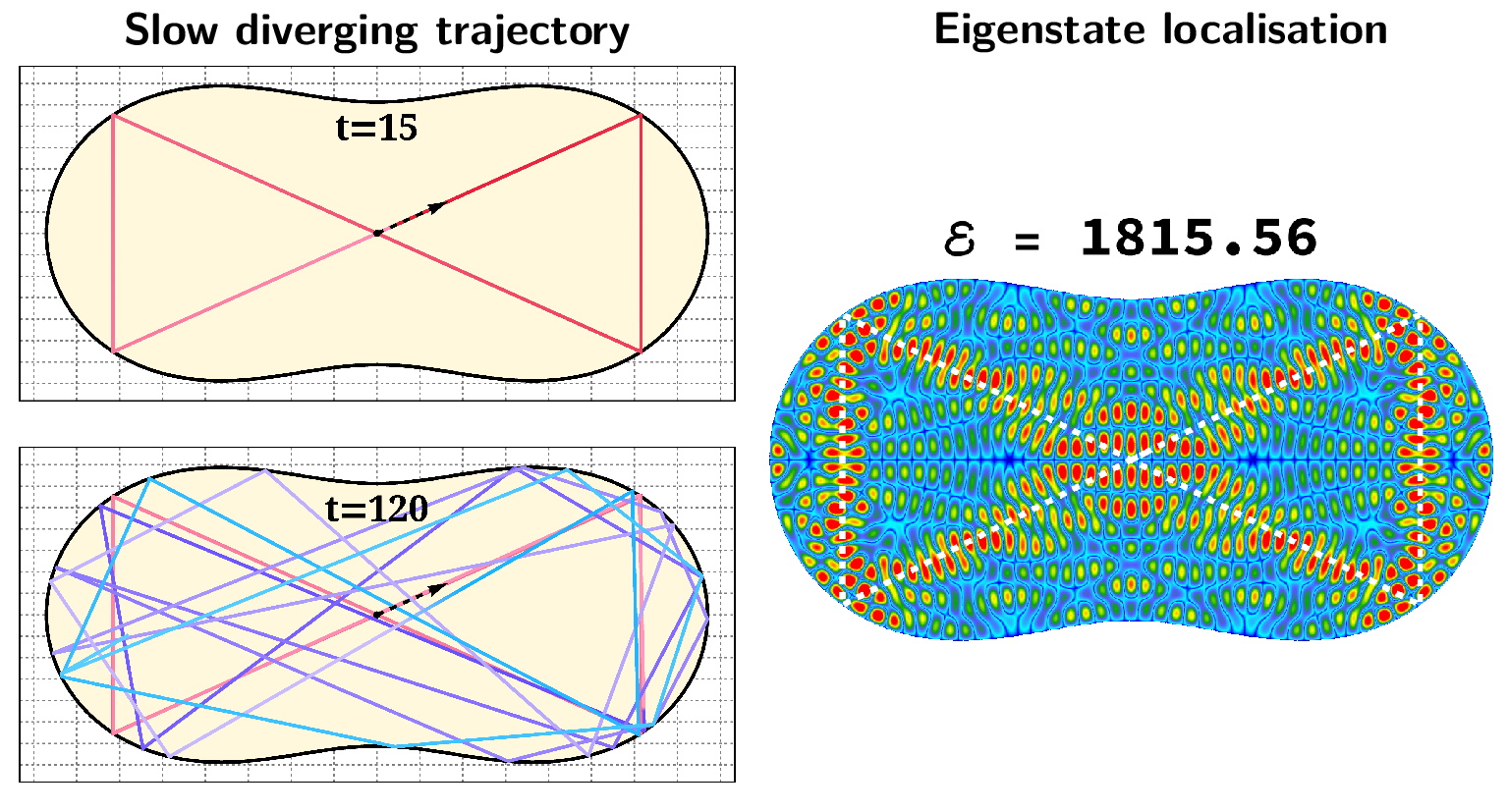}
	\caption{In the left column, a slow diverging trajectory in the peanut billiard shows almost periodic and chaotic behaviour in the early and late time, respectively. The right column shows a probability density plot producing a scar at energy $\mathcal{E}=1815.56$. It is evident that the quantum systems keep a memory of its classical IC even in the infinite time limit.}
	\label{fig9}
\end{figure}

This effect is particularly evident when the trajectories governing the classical system are chaotic. However, this does not happen in the corresponding quantum system. The quantum system retains a memory of its classical IC, even in the infinite time limit, thereby disrupting the ergodic nature of the system. This suggests that the system has an enhanced probability of remaining around its non-periodic, slow diverging trajectories. This leads to a type of localisation in the probability density of quantum observables irrespective of periodicity. These localisations are not limited to a single eigenstate. Instead, they affect a wide range of quantum eigenstates and have some structure along any trajectory as long as the rate of divergence is slow. Different names have been proposed by different authors for this kind of localisation: ``quantum trails''\cite{PhysRevLett.134.140402}, ``quantum birthmark''\cite{graf2024birthmarksergodicitybreakingquantum}. In \cref{fig9}, a visualisation on this account is provided, taking the peanut-shaped billiard as an example. Presented in the left column is a slowly diverging classical trajectory  for an IC at two different times: at $t=15$, nearly periodic behaviour and at $t=120$, chaotic behaviour. The accompanying probability density graph (on the right) shows that the eigenstate, with $\mathcal{E}=1815.56$, is sharply localised along the classical trajectory at the early time. A few other scars for slow diverging classical trajectories are provided in Appendix-\ref{apndx-A}, \cref{fig18} for both bean and peanut shaped billiards. It is possible for different eigenstates of a quantum system to exhibit scars associated with the same classical slow-diverging trajectory.

Unlike scars, super-scars\cite{PhysRevLett.100.204101, Bogomolny_2021, PhysRevResearch.4.013138, PhysRevLett.97.254102} are more structured and strongly localised that arise not from unstable/stable periodic orbits, but rather from higher symmetry or dynamical constraints in the system. They represent more extensive and stable quantum states that are confined to special regions of phase space (e.g., modes confined to certain shapes or regions of the billiard table). These states are often related to ``nongeneric''\cite{PhysRevLett.53.1515, BOGOMOLNY1988169} features of the classical system, such as symmetry or specific boundary conditions, that impose additional constraints on the wavefunctions. In these cases, the classical periodic orbit structure is not known.

The presence of both types of localisation depends on the system's geometry, symmetries, and dynamical properties, with different eigenstates reflecting different types of localisation. The quantum mechanical scars in classical chaotic systems weakly breaks ergodicity, meaning that the system does not explore all possible states equally over long times. Scars serve as vivid illustrations of how quantum mechanics suppresses chaos\cite{KeskiRahkonen2019}.

\begin{figure*}[hbt!]
	\centering
	\includegraphics[width=\linewidth]{"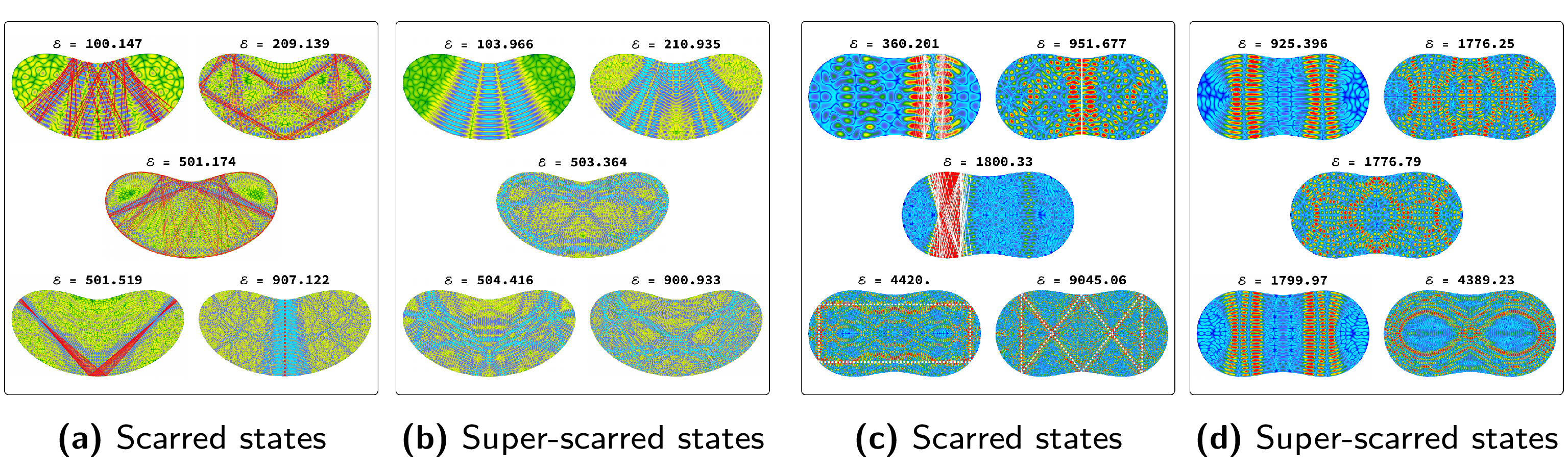"}
	\caption{The probability density distributions for eigenstates are displayed for billiards: (a \& b) the bean-shaped billiard and (c \& d) the peanut-shaped billiard. In both chaotic billiards, some eigenstates (a \& c) concentrate along specific isolated, trajectories of the corresponding classical system, highlighted by solid red and white lines. These concentrated patterns, known as quantum scars. In contrast, for the super-scarred states observed in the bean- and peanut-shaped billiards (b \& d), there is no corresponding classical orbit present.}
	\label{fig10}
\end{figure*}

\cref{fig10} shows probability density structures of quantum states in the above mentioned mixed-curvature billiard systems, highlighting regions where the density is higher. For the bean billiard, the energy of the eigenstates can be anywhere between $ 40 $ and $ 910 $, while for the peanut billiard, it ranges from $ 350 $ to $ 9056 $. The overall background represents the typical distribution of the quantum state in a chaotic system, which is usually spread out and somewhat uniform. Scarred regions, \cref{fig10}(a \& c), are the areas where the probability density is significantly higher. They appear as dark isolated lines or spots in the diagram, corresponding to the classical periodic orbits (solid red \& white lines). On the other hand, in the super-scarred states (\cref{fig10}(b \& d)), the probability densities do not align with any classical orbits, indicating a lack of classical periodic orbit influence. Compared to the bean-shaped billiard, which possesses a single symmetry axis, the peanut-shaped billiard's eigenstates demonstrate enhanced scarring owing to its two mirror symmetry axes. In contrast to non-integrable billiards, the eigenfunctions of integrable billiards are confined to specific subregions of the billiard\cite{Backer4160260}, predominantly localised around the caustics\cite{10.21468/SciPostPhys.17.5.147} found in their corresponding classical systems. 

These localised regions act like mirages in the quantum landscape, where the eigenfunction gets focused because of the underlying classical dynamics and symmetry. A comparison of scars across various energy levels reveals that they emerge only for a selected few eigenvalues. The level of scarring associated with specific eigenstates varies significantly from one state to another, although it aligns with a theoretically predicted distribution. This selective nature makes these scars especially intriguing. Scars are a concept absent in the realm of classical mechanics.

\subsection{\label{sec4B}Level spacing distribution}

The \textit{Bohigas-Giannoni-Schmit} (BGS) conjecture\cite{PhysRevLett.52.1}, states that all quantum systems (governed by autonomous Hamiltonians and ergodic behaviour) whose classical analogues are chaotic exhibit the statistical properties described by Gaussian \textit{Random Matrix Theory} (RMT)\cite{GUHR1998189} when examined in the semi-classical limit. In RMT, one studies matrices whose entries are random variables, typically drawn from specific probability distributions. These matrices can be real symmetric, complex Hermitian, or quaternionic self-dual, which respectively correspond to different symmetry classes (Dyson's three classes) and ensembles: \textit{Gaussian Orthogonal Ensemble} (GOE) statistics with \textit{Anti-Unitary Symmetry} (AUS), \textit{Gaussian Unitary Ensemble} (GUE) statistics with no AUS and \textit{Gaussian Symplectic Ensemble} (GSE) statistics with symplectic symmetry. The strength of RMT lies in its universality\footnote{Many microscopic details become irrelevant at large $\mathcal{N}$, and statistics depend only on symmetry.}; it shows that systems as diverse as atomic nuclei, chaotic billiards, and financial markets often exhibit statistical behaviours that random matrices can accurately describe. In other words, chaos within isolated quantum systems manifests as spectral correlations similar to those observed in RMT. Instead of focusing on individual eigenvalues, RMT emphasises the statistical properties of the spectrum, such as the distribution of energy levels and their correlations.

The NNSD serve as a primary indicator of quantum chaos among many statistical measures. In this context, we define level spacing as the energy gap between two neighbouring levels in the unfolded spectrum, denoted by $ s_{i}=\tilde{\mathcal{E}}_{i+1}-\tilde{\mathcal{E}}_{i} $. The unfolding procedure ensures that the mean level spacing is normalised to unity. The \textit{level spacing distribution}, $ P(s) $, takes centre stage, providing degree of level repulsion. A dramatic insight in the field of quantum chaos is summarised by the universality conjectures for $P(s)$:
\begin{equation}{\label{eq16}}
	P(s)= 
	\begin{cases}
		\mathrm{e}^{-s} &\text{Poissonian},\\
		\dfrac{\pi s}{2} \mathrm{e}^{\dfrac{-\pi s^2}{4}} & \text{GOE,}\\
		\dfrac{32 s^2 \mathrm{e}^{-\dfrac{4 s^2}{\pi }} }{\pi ^2} & \text{GUE}
	\end{cases}
\end{equation}

\begin{figure}[hbt!]
\centering
\includegraphics[width=\linewidth]{"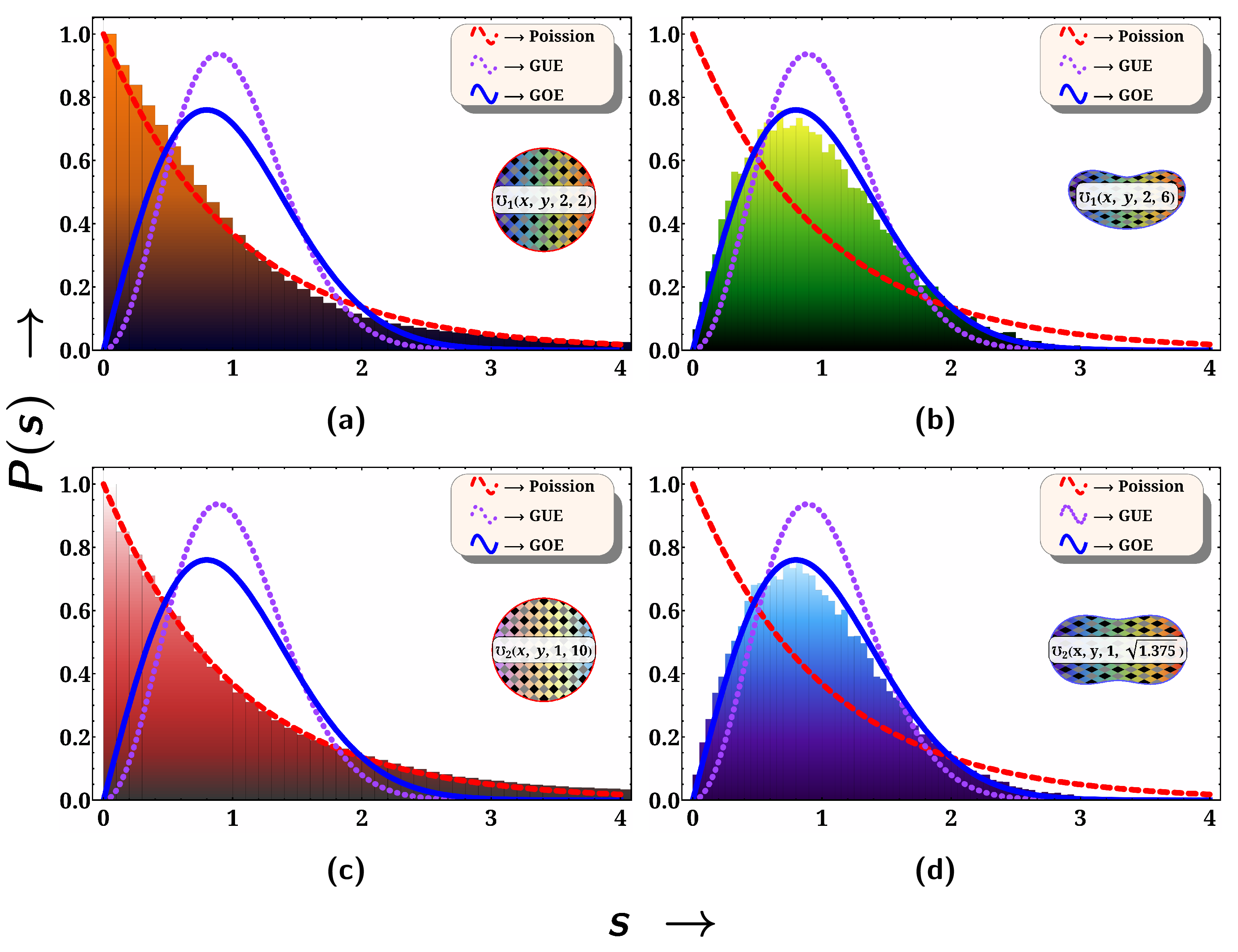"}
\caption{\label{fig11}Comparison of NNSD $P(s)$ for various quantum systems with standard RMT predictions. The histograms represent the level spacing distributions for different billiard geometries: (a) Circle, (b) Bean (c) Ellipse, and (d) Peanut billiards. The theoretical curves correspond to Poisson (red dashed line), GOE (blue line), and GUE (purple dotted line) statistics. The observed behaviours align closely with Poisson statistics for circle and ellipse billiards and GOE predictions for bean and peanut billiards.}
\end{figure}

``Level repulsion''\cite{Haake2018_ch3} (i.e., $ P(s) \to 0 $) is characteristic of chaotic systems, in contrast to integrable systems, which display ``level attraction'' (i.e., $ P(s) \to 1 $). The \textit{Wigner-Dyson} (WD) distribution\cite{PhysRevLett.52.1, casati1980connection} offers a good approximation of these characteristics for chaotic systems. WD or GOE indicate towards the presence of avoided crossings. Meanwhile, for generic integrable classical systems, the NNSD follows Poisson distribution\cite{berry1977level}. Poisson distribution imply energy levels are uncorrelated, and level crossings are allowed\cite{doi:10.1098/rspa.1977.0140}. In essence, the observation of a Poisson or GOE distribution in the level spacing serves as a distinct characteristic of a quantum integrable or chaotic system, respectively.

In \cref{fig11}, the circular and elliptical billiards (\cref{fig11}(a \& c)) exhibit Poisson distributions, which aligns with their classical nature. This agreement is excellent, as it implies that the energy levels have no correlation. Meanwhile, for the bean- and peanut-shaped billiards ({\cref{fig11}}(b \& d)), we observe a distribution resembling close to GOE, confirming the predominant chaotic nature of the system. Conversely, systems with mixed dynamics show an intermediate statistics between Poisson and GOE distributions. These quantum signatures are consistent with the classical behaviour revealed by Poincaré sections. To verify the convergence of these results, we ran numerical evaluations for $\mathcal{N} = 2000$, $ 5000$, $ 20000$, and $ 40000$, yielding consistent results for each of the billiards.

Whenever the agreement of one of the distributions $P(s)$ with data or numerical calculations is tested, it is helpful to use the \textit{Cumulative Level-Spacing Distribution} (CLSD)\cite{mehta2004random, Haake2010} given by $I(s)=\int_{0}^{s}P(s^{\prime})ds^{\prime}$. It is independent of binning effects\footnote{Data binning or bucketing is a data preprocessing method used to minimize the effects of small observation errors}. In other words, it is much smoother and less sensitive to statistical fluctuations (noise) than the raw NNSD, especially for small or moderately sized datasets.

\begin{figure}[hbt!]
\centering
\includegraphics[width=\linewidth]{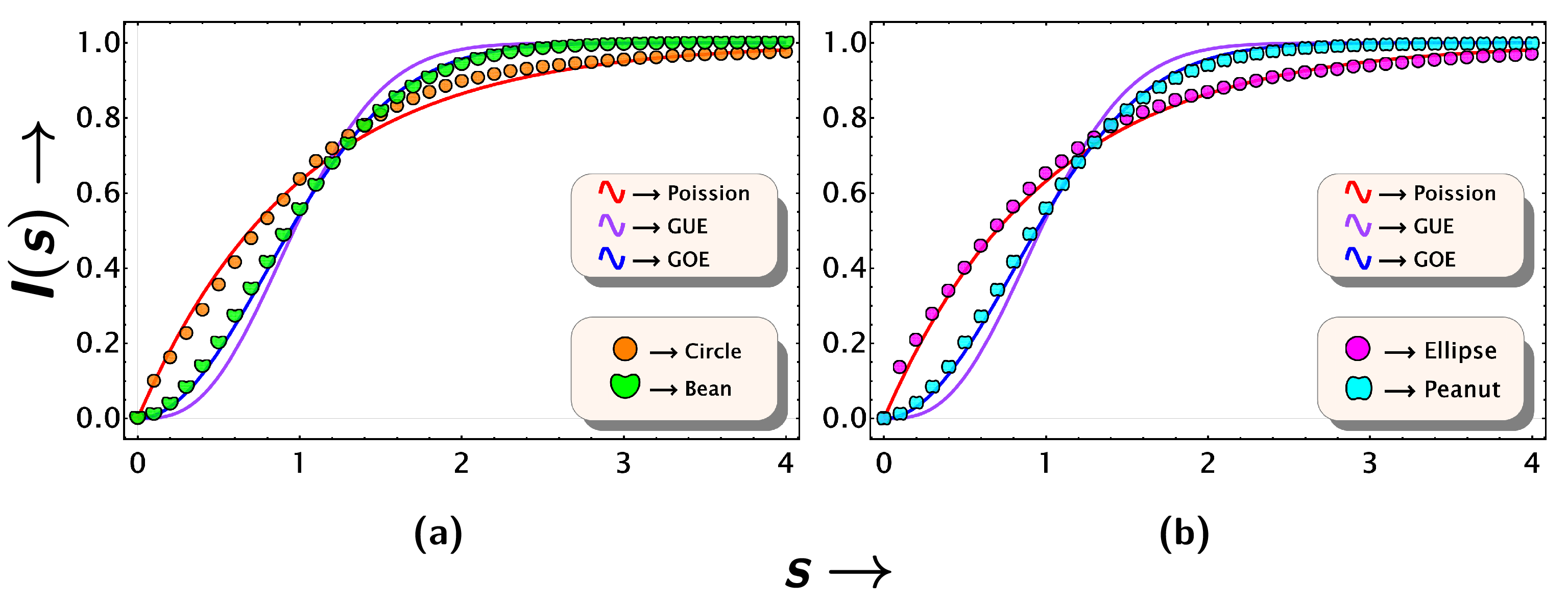}
\caption{The CLSD, $I(S)$, for the four billiards. Circular and elliptical billiard follows Poission distribution, while Bean and Peanut billiard aligns closely with GOE. Deviation from the empirical curve signals mixed dynamics or finite-size effects.}
\label{fig12}
\end{figure}

\cref{fig12} shows the variation of $I(s)$ for the four billiards. For circular and elliptical billiard, $I(s)$ exhibits a rapid rise at small $s$. It is in close agreement with the empirical Poisson curve and supports the systems integrability. While for bean and peanut, $I(s)$ follows a sigmoidal shape, with clear suppression at $s\sim 0$. Their close proximity to the GOE curve indicates dominant chaotic behaviour. 

\subsubsection{\label{sec4B1}Level spacing ratio}
To strengthen the validity of our results, we employed another commonly used short-range statistical measure, the LSR\cite{PhysRevLett.110.084101, PhysRevResearch.4.013138}. Oganesyan and Huse\cite{PhysRevB.75.155111} in $2007$ introduced the distribution of the ratios $ r_{i} $ defined by
\begin{equation}\label{eq17}
\expval{\widetilde{r}}_{i} = \dfrac{\min(s_{i}, s_{i-1} )}{\max(s_{i}, s_{i-1} )} = min(r_{i},\dfrac{1}{r_{i}} ),
\end{equation}
where 
\begin{equation}\label{eq18}
r_{i}=\dfrac{s_{i}}{s_{i-1}}
\end{equation}
is the the ratio of two consecutive level spacings. The LSR stands out as a convenient and reliable measure of spectral statistics because it avoids the need to unfold the spectrum and is unaffected by local density of states variations. This measure offers a statistical snapshot of the underlying dynamics responsible for the system's behaviour.

In particular, the mean $\widetilde{r}$ fluctuates around the following values, depending on the distribution of adjacent level spacing in the spectrum being studied \cite{PhysRevLett.110.084101}.
\begin{equation}{\label{eq19}}
\expval{\widetilde{r}}=
\begin{cases}
	2 \ln(2) - 1 \approx 0.38629 &\text{Poissonian},\\
	4-2 \sqrt{3} \quad\approx 0.5307& \text{GOE,}\\
	\frac{2 \sqrt{3}}{\pi}- \frac{1}{2} \quad\approx	0.5996 &\text{GUE}
\end{cases}
\end{equation}
Meanwhile, the average value of $r$ for Poisson is $\expval{r}_{P}=\infty$, whereas for GOE, it is $\expval{r}_{GOE} = 1.75$ \cite{chavda2015distribution}. The mean $\tilde{r}$ and $r$ for our four billiards are shown in \cref{table3}. Although unfolding is not strictly required, comparing statistics in terms of unfolded energies proves most convenient.

For the bean- and peanut-shaped billiards, the mean LSR values tend to fluctuate around $\expval{r}_{GOE}$, indicating chaotic behaviour. In contrast, for the circular and elliptical billiards, the LSR values hover around $\expval{r}_{P}$, consistent with their integrable nature.

{\renewcommand{\arraystretch}{1.5}
	\begin{table}[hbt!]
		\centering
		\caption{\label{table3}The mean $\widetilde{r}$ and $r$ for bean-shaped and Cassini ovals billiards.}
		\begin{tabular}{ rcl} 
			\toprule[0.07cm]
			\midrule
			\texttt{Billiards} & $\expval{\widetilde{r}}$ & $\expval{r}$ \\
			\hline	
			\texttt{\textit{Circle:}} &  $\sim 0.395968 $ & $\sim 39895.1 ~(\approx\infty)$\\
			\texttt{\textit{Bean:}} &  $ \sim 0.518245 $ & $ \sim 1.93933$ \\
			\texttt{\textit{Ellipse:}} &  $ \sim 
			0.382986 $ &  $ \sim 16758.6 ~(\approx\infty)$ \\
			\texttt{\textit{Peanut:}} &  $ \sim 0.520381 $ & $\sim 1.9771$\\
			\bottomrule[0.07cm]	
		\end{tabular}
	\end{table}
}

\subsubsection{\label{sec4B2}Spectral staircase function}
Another statistical measure is the spectral staircase function (or the level counting function), $N(\mathcal{E})$, which counts number os eigenstates with energies less than or equal to a specified energy value $\mathcal{E}$ \cite{PhysRevResearch.4.013138, PhysRevE.52.2463}. Formally it is defined as, $N(\mathcal{E}):= \# \{\mathcal{N} \in \mathbb{N} ~|~ \mathcal{E_{N}}\le\mathcal{E}\}$. For billiard systems, the average $N(\mathcal{E})$ follows  generalised Weyl's law\cite{Dowker_1978, backer2002behaviour, mueller2007weylslawtheoryautomorphic}:
\begin{equation}\label{eq20}
N_{Weyl}(\mathcal{E}) = \dfrac{1}{4 \pi} \bigg(\abs{\varOmega}~\mathcal{E}^{\frac{\mathtt{d}}{2}} + \abs{\dd{\varOmega}}~\mathcal{E}^{\frac{\mathtt{d}-1}{2}}\bigg) + \mathcal{R}(\mathcal{E}), 
\end{equation}
where $\mathtt{d}$ is the system's spatial dimension, $\abs{\varOmega}$ denotes the area of the billiard, $\abs{\dd{\varOmega}}$ is the length of the boundary provided in \cref{table4}. In smooth domains, where the boundary is sufficiently regular, $\mathcal{R}(\mathcal{E})$ decays rapidly. However, for domains with singularities (e.g., corners, edges or cusps), $\mathcal{R}(\mathcal{E})$ is more intricate and may decay more slowly, though it remains bounded by $\order{\mathcal{E}^{\frac{\mathtt{d}-1}{2}}}$. Thus, the spectral staircase naturally separates into a leading smooth part and a fluctuating remainder.

{	\renewcommand{\arraystretch}{1.5}
	\begin{table}[hbt!]
		\centering
		\caption{\label{table4} Area and boundary length of the Billiards}
		\begin{tabular}{ rcc } 
			\toprule[0.07cm]
			\midrule
			\texttt{Billiards} & \texttt{Area $\big(\abs{\varOmega}\big)$} & \texttt{Perimeter $\big(\abs{\dd{\varOmega}}\big)$} \\
			\midrule
			\texttt{\textit{Circle:}}  & $4 \pi$ &  $4 \pi$ \\
			\texttt{\textit{Bean:}}  & $31.4159 $ & $22.2411$ \\
			\texttt{\textit{Ellipse:}}  & $314.151$ & $62.8322$ \\
			\texttt{\textit{Peanut:}}  & $3.67382$ & $7.68519$ \\
			\bottomrule[0.07cm]	
		\end{tabular}
	\end{table}
}

\begin{figure}[hbt!]
	\centering
	\includegraphics[width=\linewidth]{"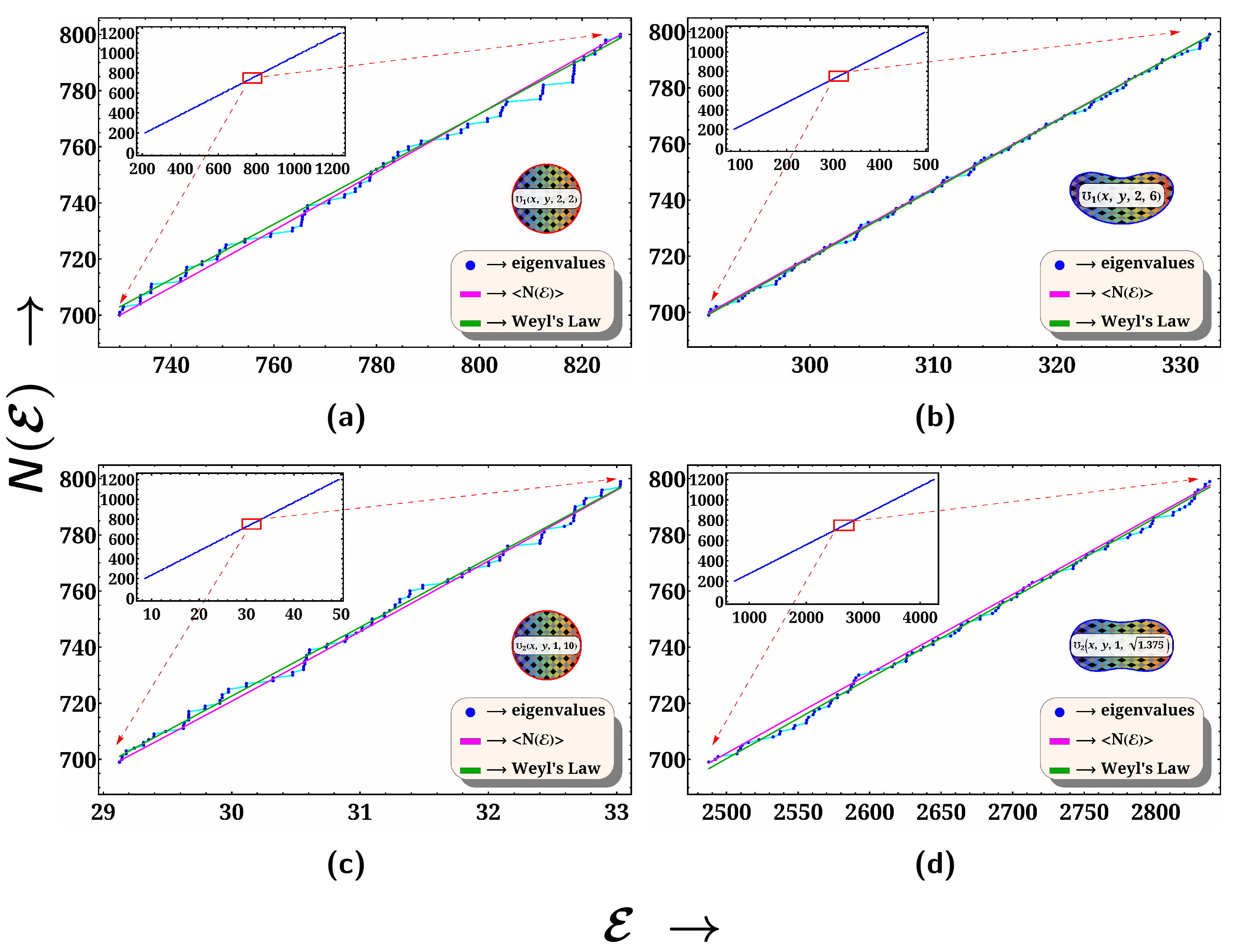"}
	\caption{The plot displays the spectral staircase function $N(\mathcal{E})$ for (a)Circular, (b) Bean, (c) Ellipse, and (d) Peanut billiards, with its characteristic steps corresponding to discrete eigenvalues. While the red line provides the overall trend, the green line corresponds to Weyl's law. Both the lines are seen to describe the average behaviour of $N(\mathcal{E})$, reasonably well. $\mathcal{N} \in [700, 800\label{key}]$}.
	\label{fig13}
\end{figure} 

In integrable systems, $N(\mathcal{E})$ grows smoothly with small, regular fluctuations. This shows that the level spacing is uniform. On the other hand, chaotic systems have irregular and complex fluctuations in $N(\mathcal{E})$, indicating non-uniform level spacing. These irregular fluctuations are often connected to phenomena like quantum interference and the scarring of wavefunctions along classical unstable periodic orbits. In \cref{fig13}, we present the spectral staircase function, which adequately highlights the contrasting characteristics of integrable and chaotic behaviour in our billiard models. The spectral staircase functions for circular (\cref{fig13}(a)) and elliptical (\cref{fig13}(c)) billiards exhibit nearly identical growth patterns, with small fluctuating steps arranged in a relatively regular manner. In contrast, for bean-shaped (\cref{fig13}(b)) and peanut-shaped (\cref{fig13}(d)) billiards, the staircase function fluctuates irregularly with uneven intervals, providing a distinct signature of chaos.

The outcomes derived from diagnostic measures such as $P(s)$, $I(s)$, LSR, and $N(\mathcal{E})$, collectively indicate that the statistical characteristics of the quantum energy spectrum faithfully mirror the underlying classical dynamics. In these mixed-curvature systems, $P(s)$ tends toward WD statistics, indicating level repulsion. However, it still keeps features of uncorrelated spectra, reflecting the coexistence of regular islands embedded within a predominantly chaotic phase space. The $I(s)$, being less sensitive to local fluctuations, follows the theoretical predictions of pure ensembles and thus acts as a smoother indicator of the underlying dynamics. The LSR for bean and peanut, with values close to the mean ratios, also reflects this dominant chaotic behaviour. $N(\mathcal{E})$ shows irregular fluctuations superimposed on the average Weyl's law behaviour. In summary, these hybrid structures (bean and peanut) exhibit chaos dominant quantum spectral statistics.

\subsection{\label{sec4C}Spectral complexity}

Complexity is a characteristic of a quantum state\cite{https://doi.org/10.1002/prop.201500095, https://doi.org/10.1002/prop.201500092}. Despite its importance, the information embedded within it is often disregarded because its intricacies seldom translate into the local observable properties of the system\cite{RevModPhys.82.277}. Complexity in quantum systems is extensive, meaning it scales with the number of active degrees of freedom. The concept of SC, originally formulated in the context of black hole studies\cite{iliesiu2022volume, Iliesiu2022}, has since found applications beyond this, including the diagnosis of chaos\cite{PhysRevD.109.046017, Balasubramanian2025, Camargo2024}. SC incorporates thermal averaging, meaning it reflects how energy levels are populated at thermal equilibrium.

In this section, we use SC to probe chaotic behaviour within billiard systems. Here, the focus is on the time-scale at which the saturation occurs in the growth rate of SC. In more formal terms, for quantum systems possessing a discrete, non-degenerate energy spectrum, the SC, denoted as $C_{s}(t)$, is defined as\cite{PhysRevD.109.046017, Camargo2024}:
\begin{widetext}
\begin{equation}{\label{eq21}}
C_{s}(t):=\dfrac{1}{D~ Z(2 \beta)} \sum_{\mathcal{M} \neq \mathcal{N}} \Bigg(\dfrac{\sin(\dfrac{t(\mathcal{E}_{\mathcal{M}}-\mathcal{E}_{\mathcal{N}})}{2})}{\dfrac{(\mathcal{E}_{\mathcal{M}}-\mathcal{E}_{\mathcal{N}})}{2}}\Bigg)^{2}\mathrm{e}^{-\beta (\mathcal{E}_{\mathcal{M}}+\mathcal{E}_{\mathcal{N}})}
\end{equation}
\end{widetext}
where $D$ is the dimension of the Hilbert space $\boldmath{\mathbb{H}}$, $Z(\beta)$ is the thermal partition function, $\beta = \frac{1}{T}$ is the inverse temperature, and \{$\mathcal{E}_{1},\mathcal{E}_{2},\mathcal{E}_{3},\ldots,\mathcal{E}_{\mathcal{N}}$\} are the discrete energy eigenvalues. Here, temperature is introduced as a formal parameter for thermal averaging over the energy spectrum. In the limit $T \to \infty$, the thermal weights become uniform, giving equal statistical weight to all energy levels in the spectrum. This approach effectively transforms spectral complexity into a global measure, consistent with the NNSD predicted by RMT, which captures universal level statistics. At finite temperatures, however, the averaging becomes energy-selective, emphasising contributions from eigenstates within a narrower energy window, thereby providing temperature-dependent insight into spectral correlations. Authors\cite{erdmenger2023universal, PhysRevD.109.046017, Camargo2024} have shown that SC plateaus to different values across different time scales based on the integrability of the system. 

SC typically exhibits three temporal regimes\cite{Balasubramanian2025} for the rescaled spectra with a bounded range in the leading large-$D$ limit. These three different regimes of SC reveal how chaos emerges, how it spreads, and how it reaches equilibrium. At early times ($t\ll 1$), SC grows quadratically ($\sim t^{2}$), reflecting uncorrelated level contributions. This behaviour is largely independent of geometry and reflects universal short-time dynamics as predicted by perturbative expansions. For $1 \leq t \leq D$, the growth shifts to a linear pattern ($2 \pi t$), indicating the start of spectral correlations. In this regime, the system enters the chaotic mixing stage characterised by decaying correlations, diffusive operator growth. Here, spectral properties resemble RMT predictions, indicating that the system is now exploring a substantial portion of its spectrum, not just immediate neighbours. From an information-theoretic point-of-view, linear growth signifies that local information is becoming globally distributed (non-locally encoded) across the system at the fastest possible rate. At late times ($D \ll t$), SC becomes highly sensitive to spectral statistics. The operator has fully thermalised and has explored the available Hilbert space. It has reached a saturation plateau at ``Heisenberg time\footnote{Heisenberg time $t_{H}=2 \pi \rho(\mathcal{E})$, where $\rho(\mathcal{E})$ $\Big(=\pdv{N}{\mathcal{E}}\Big)$ is the mean density of states and under normalisation condition, $\int \dd{\mathcal{E}}\rho(\mathcal{E})=D$. Since $\rho(\mathcal{E})$ is systems specific, $t_{H}$ is an explicit, geometry-dependent time scale for billiards.}'' ($t_{H}$) following a logarithmic trend. The saturation is a direct consequence of the finite dimension of the Hilbert space. The maximum value of $C_{s}$ scales with system size. Specifically, for integrable billiards, the saturation value of SC is significantly higher, by several orders of magnitude, compared to non-integrable billiard systems. This contrast highlights how SC stabilisation depends on whether the underlying dynamics are integrable or chaotic. At late times, SC shows the following behaviours\cite{Balasubramanian2025} depending on the integrability and degeneracy of the system,
\begin{equation}{\label{eq22}}
C_{s}(t) \propto
\begin{cases}
	t &\text{Poissonian},\\
	\ln(t) & \text{GOE,}\\
	t^{2}&\text{Degenerate Spectra}
\end{cases}
\end{equation}

For quantum systems with non-degenerate energy spectra, the SC is constrained as $ \sin^{2}\Big(\frac{t(\mathcal{E}_{\mathcal{M}}-\mathcal{E}_{\mathcal{N}})}{2}\Big) \leq 1$ in \cref{eq21}\cite{Balasubramanian2025}. This leads to an upper bound on $ C_{s}(t) $ given by $ \Big(\frac{2}{\Delta \mathcal{E}_{min}}\Big)^{2} $, where $ \Delta \mathcal{E}_{min} $ is the minimum energy level spacing. Here, $ \Delta E_{\min} $ scales to $ \frac{1}{D} $, since we have unfolded the eigenvalue spectrum to have a width of $ 1 $. And at infinite temperature (where $ \beta = 0 $), the saturation value depends solely on the energy difference $ \Delta \mathcal{E}:=\mathcal{E}_{\mathcal{M}}-\mathcal{E}_{\mathcal{N}} $, not on $ \mathcal{E}_{\mathcal{M}}+\mathcal{E}_{\mathcal{N}} $ (See Appendix-\ref{apndx-B}). Thus, the saturation value of complexity can reach greater heights when the difference in energy is at its smallest. Therefore, $\Delta \mathcal{E}_{min}$ represents a key inverse time scale that determines the saturation of SC\footnote{The sine-squared factor acts like an energy resolution filter of width $ \Delta \mathcal{E} \sim \frac{1}{t} $. This behaviour is a direct consequence of the time-energy uncertainty principle. Increasing $t$ improves energy resolution: only eigenvalue pairs with small level spacing $ \abs{\Delta \mathcal{E}_{MN}}\lesssim \frac{1}{t} $ contribute. Thus, scanning $ t $ is equivalent to scanning energy differences: small $ t $ sees numerous pairs incoherently (giving the $ t^{2} $ start); larger $ t $ picks out fine structure in the level spacing statistics.}. 

\begin{figure}[hbt!]
\centering
\includegraphics[width=\linewidth]{"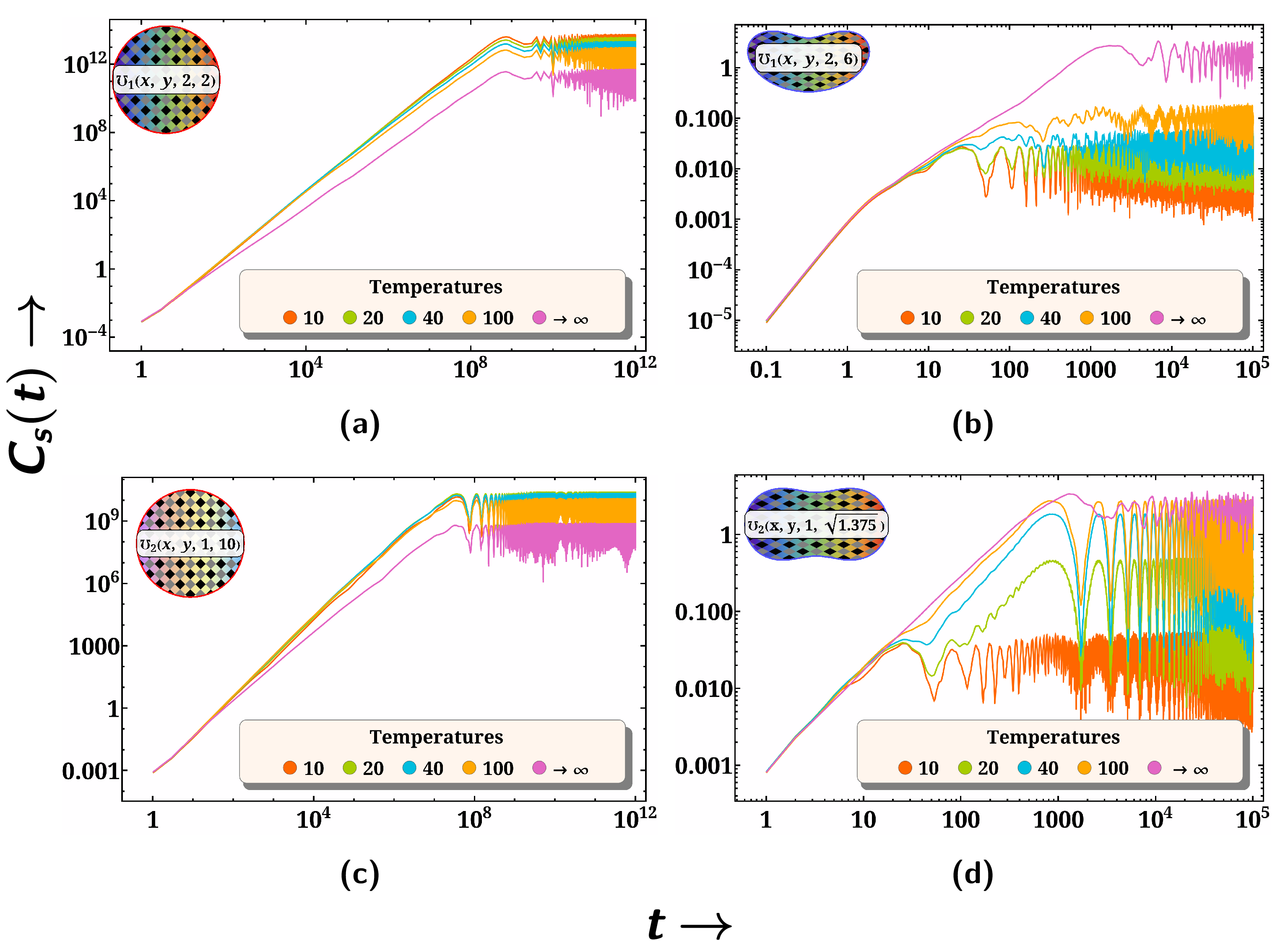"}
\caption{\label{fig14}Temperature dependence of spectral complexity at $ T = 10, ~20, ~40, ~100 $ and at $ \to \infty $ for (a) Circle, (b) Bean (c) Ellipse, and (d) Peanut billiards. As expected, Bean and Peanut billiards exhibit logarithmic late-time growth consistent with the GOE. Conversely, Circular and elliptical billiards show linear late-time growth, mirroring the expected behaviour of a Poisson spectrum.}
\end{figure}

Since $ \Delta \mathcal{E}_{min} \geq0 $, the $ C_{s}(t) $ reaches saturation significantly earlier in chaotic systems than in integrable ones, where $ \Delta \mathcal{E}_{min} $ approaches zero (in the semi-classical limit). This early saturation in chaotic systems suggests a faster transition to complexity, reflecting the underlying chaotic nature compared to the gradual, regular build-up in integrable systems. This dichotomy makes it a powerful tool to distinguish chaos from regularity. This trend aligns with observations in bean- and peanut-shaped billiards, where the onset of chaos leads to an apparent early saturation of SC growth. In contrast, integrable billiards (circle and ellipse) maintain a persistent rise in complexity (\cref{fig14}).

To make the Hilbert space of an infinite-dimensional quantum system manageable for practical calculations, we truncate it to a finite dimension, specifically setting $ D = 1000 $. This choice of $ D $ is specifically suitable considering, computational time, convergence of $ C_{s}(t) $, and resolution between different temperatures. For a quantum system with $ D- $dimensional Hilbert space and non-degenerate energy spectrum, the number of distinct lowest energy eigenvalues $ \mathcal{N} $ can match the truncated Hilbert space dimension $ D $\cite{Aleman2012}. In integrable systems, $ \mathcal{N} $ acts as a resolution parameter: higher $ \mathcal{N} $ values improve the distinguishability of $ C_s(t) $ at different temperatures. This is because SC, in its definition, has input from the full spectrum of the system. For computing SC, the essential component is the energy spectrum, which we obtain numerically following \cref{eq15}.

Here, in \cref{fig14}, we observe a clear distinction in the late-time behaviour of $ C_s(t) $ between different billiard shapes. Specifically, for integrable billiards with circular (\cref{fig14}(a)) and elliptical (\cref{fig14}(c)) shapes, SC saturates at a noticeably later timescale than for non-integrable billiards with bean (\cref{fig14}(b)) and peanut (\cref{fig14}(d)) shapes. In chaotic (non-integrable) systems, level repulsion, the tendency for energy levels to avoid crossing or clustering, prevents the energy difference $\Delta \mathcal{E}_{min}$ from approaching zero. As a result, the saturation timescales for SC in chaotic systems are significantly shorter than those observed in integrable systems, revealing a disparity of several orders of magnitude.

While prior studies have been restricted to the infinite temperature regime\textemdash where Boltzmann weights are uniform\textemdash we extend the analysis of SC to both finite and infinite temperatures. This broader perspective allows us to uncover how thermal fluctuations modulate the growth and saturation of SC, thereby revealing the temperature dependence of quantum chaotic behaviour. The saturation time of $C_s(t)$ in circular and elliptical billiards shows a negligible dependence on temperature, remaining nearly constant. Conversely, in bean- and peanut-shaped billiards, lower temperatures hinder the growth of complexity, making these systems less complex. Higher temperatures, however, lead to significantly greater growth of SC, which saturates at a much higher level and over a longer period. This consistency suggests a fundamental difference in how integrable and chaotic systems respond to temperature fluctuations in terms of complexity growth. Notice that, as the temperature decreases, SC becomes increasingly sensitive to local features of the spectrum within the energy window where the Boltzmann weight $\mathrm{e}^{-\beta \mathcal{E}}$ remains significant. At absolute zero ($\beta \to \infty$), the SC primarily investigates the ground state and nearby excited states because high-energy states are suppressed by thermal effects.

\begin{figure}[hbt!]
	\centering
	\includegraphics[width=\linewidth]{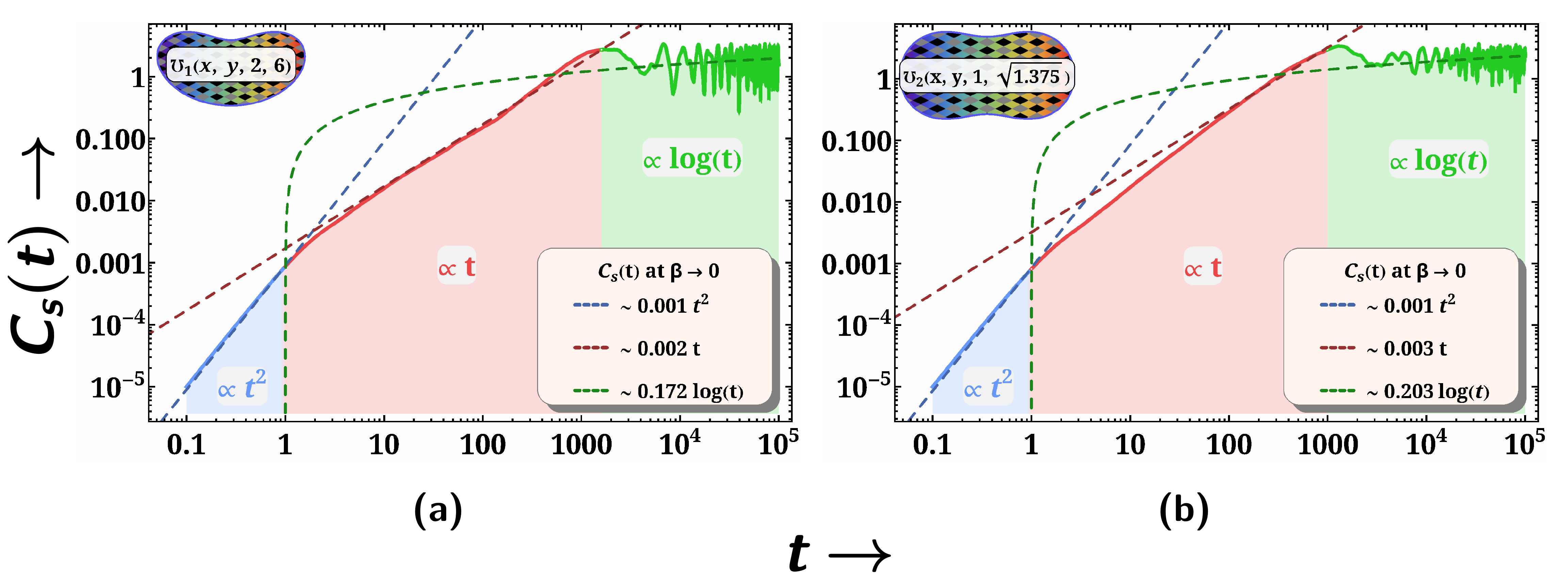}
	\caption{$C_{s}(t)$ vs $t$ for the chaotic billiards in the infinite temperature limit (i.e., $\beta \to 0$) for (a) bean-shaped billiard and (b) peanut-shaped billiard. Both sub-figures display the characteristic early-time quadratic rise (blue, $\propto t^{2}$), followed by an intermediate linear regime (red, $\propto t$), and finally a long-time logarithmic growth (green, $\propto \log(t)$) in the $\beta \rightarrow 0$ limit. Best-fit prefactors for the limiting behaviour\textemdash quadratic, linear, and logarithmic\textemdash are shown in the dashed lines. While both systems share a similar level of global chaos, the peanut billiard shows a significantly slower transition from quadratic to linear behaviour.}
	\label{fig15}
\end{figure}

\cref{fig15} shows the growth of SC over time for the two chaotic billiard geometries. Both systems initially for $t<1$ display quadratic growth, indicating uncorrelated level contributions. For $1<t<D$, spectral correlations develop slowly but steadily with a linear trend, reflecting the start of spectral correlations. This growth is slower in peanut-shaped billiard as compared to bean-shape. At late times ($t>>D$), following \cref{eq22}, $C_s(t)$ saturates into a plateau. This follows logarithmic trend with slopes $\sim 0.172$ and $\sim 0.203$ for bean- and peanut-shaped billiards, respectively, marking the $t_{H}$ beyond which additional complexity no longer accumulates. The persistent oscillations around the saturation level suggest residual correlations or finite-size effects in the spectrum. This behaviour is a hallmark of quantum chaotic systems and agrees with predictions from random matrix theory for the late-time saturation of spectral statistics.

Although both the chaotic systems exhibit the same qualitative structure, their quantitative crossover times differ. Because the peanut geometry supports more pronounced trapping regions, the quantum dynamics retains more memory of the initial configuration for a longer time (resulting in more scarred states). This memory manifests as an extended precedence of non-universal, system-specific correlations. As a result, the crossover from $C_s(t)\propto t^2 \longrightarrow C_s(t)\propto t$ is slower. The bean billiard, lacking such a constriction, reaches global mixing more quickly. The transition from quadratic to linear growth marks the onset of chaos, the slope of the linear growth quantifies its strength, and the saturation reveals the finite size of the ``quantum universe'' the system inhabits.

$C_{s}(t)$ quantifies complexity by summing the squared sine-weighted contributions of level spacings ($\Delta \mathcal{E}$), and its logarithmic growth matches the ``ramp'' regime of the GOE \textit{Spectral Form Factor} (SFF)\cite{Haake2010ch10}, an indicator of chaotic dynamics. Eventually, $C_{s}(t)$ saturates, corresponding to the GOE ``plateau'', where eigenvalues are fully correlated within the spectral window. Crucially, earlier saturation indicates a faster onset of level rigidity, which is typically associated with more chaotic quantum spectra. Although the peanut shows a steeper logarithmic slope, its early saturation time closely matches that of the bean-shaped billiard. Thus, both indicates comparable build-up of long-range correlations and strong level repulsion, which is consistent with their classical chaotic dynamics. This behaviour is in contrast to Poissonian (integrable) spectra, where SC would remain nearly flat at late times, signalling the absence of level repulsion and long-range correlations. In summery, SC unifies time dynamics, thermal effects, and spectral rigidity into a single framework, offering a sharper probe of quantum chaos than static metrics.

\subsection{\label{sec4D}Out-of-time-order correlator}
The OTOC is a key diagnostic tool in quantum chaos\cite{Hashimoto2017, GarcaMata2023, Das2025} and information scrambling\cite{PhysRevLett.126.030601, Das2025}. The relation between the OTOC and chaos can be easily seen in the semiclassical limit\footnote{In the semiclassical limit, $ \comm{}{}_{comm}\to i\hbar\{,\}_{poisson}\implies C_{\beta}(t)\to \hbar^{2}\qty(\pdv{x(t)}{x(0)})^2$. This gives the sensitive dependence on ICs, the classical diagnostic of the butterfly effect (i.e., $\delta x(t)\sim\delta x(0)\mathrm{e}^{\lambda_L t}$, where $\lambda_L$ is the Lyapunov exponent).}. In quantum mechanics, it represents the growth of non-commuting operators describing the unequal time commutation relation. Mathematically, this can be expressed as a commutation relation involving two operators at distinct points in time. A $4-$point thermal OTOC\cite{Hashimoto2017, sym13010044} is defined as:
\begin{eqnarray}{\label{eq23}}
C_{\beta}(t)&\coloneq& \frac{1}{Z(\beta)} \sum_{\mathcal{N}} c_{m}(t)\mathrm{e}^{-\beta\mathcal{E}_{\mathcal{N}}},\\ \nonumber
\text{where,}~ c_{m}(t) &\equiv& -\expval{\comm{\hat{x}(t)}{\hat{p}}^{2}}{\mathcal{N}}
\end{eqnarray}
We refer $ c_{m}(t) $, in \cref{eq23}, for a fixed energy eigenstate as \textit{Microcanonical OTOC}. Here, $\hat{x}$ and $\hat{p}$ are Hermitian operators and $\hat{x}(t)=\mathrm{e}^{i \mathcal{H}t}\hat{x}\mathrm{e}^{-i \mathcal{H}t}$. The explicit form of it is borrowed from \cite{sym13010044} as given below
\begin{widetext}
	\begin{eqnarray}{\label{eq24}}
c_{m}(t)&=& \frac{1}{4} \sum _{\mathcal{K},\mathcal{L},\mathcal{R}} x_{\mathcal{NL}}~ x_{\mathcal{LK}}~ x_{\mathcal{RN}}~ x_{\mathcal{KR}} \Big( \Delta\mathcal{E}_{\mathcal{RK}}~\Delta\mathcal{E}_{\mathcal{LK}} \mathrm{e}^{i t \Delta\mathcal{E}_{\mathcal{RL}}} +\Delta\mathcal{E}_{\mathcal{NR}}~\Delta\mathcal{E}_{\mathcal{NL}} \mathrm{e}^{-i t \Delta\mathcal{E}_{\mathcal{RL}}} \nonumber\\
		&&
-\Delta\mathcal{E}_{\mathcal{RK}}~\Delta\mathcal{E}_{\mathcal{NL}} \mathrm{e}^{i t\big(\Delta\mathcal{E}_{\mathcal{RN}}+\Delta\mathcal{E}_{\mathcal{LK}}\big)} - \Delta\mathcal{E}_{\mathcal{NR}}~\Delta\mathcal{E}_{\mathcal{LK}} \mathrm{e}^{-i t\big(\Delta\mathcal{E}_{\mathcal{RN}}+\Delta\mathcal{E}_{\mathcal{LK}}\big)}\Big) 
	\end{eqnarray}
\end{widetext}
Here, $\Delta\mathcal{E}_{\mathcal{NM}} = \mathcal{E}_{\mathcal{N}}-\mathcal{E}_{\mathcal{M}}$, $x_{\mathcal{NM}} = \matrixel{\mathcal{N}}{\hat{x}}{\mathcal{M}}$ and $p_{\mathcal{NM}} = \matrixel{\mathcal{N}}{\hat{p}}{\mathcal{M}} =\frac{i}{2}\Delta\mathcal{E}_{\mathcal{NM}}x_{\mathcal{NM}} $. If $\hat{x}(t)$ and $\hat{p}$ commute, then the OTOC vanishes. Therefore, non-zero OTOC directly reflects their non-commutativity, thus reflecting sensitivity to ICs. Apart from the well-known exponential growth of the OTOC, which is widely regarded as a hallmark of quantum chaos, other growth patterns such as periodic, aperiodic, or randomly fluctuating behaviour typically signal the absence of chaotic dynamics\cite{sym13010044}, reflecting integrability or near-integrability. In chaotic systems, OTOC grows exponentially up to the scrambling or Ehrenfest time\cite{Sinha_2024, PhysRevLett.118.086801}. After that, because of finite size effects in a bounded system (such as a billiard), OTOCs will fluctuate around a constant saturation value in the later times rather than grow indefinitely. The amplitude of the fluctuations decreases as the chaos parameter becomes larger, approaching small quasi-random fluctuations in the fully chaotic regime\cite{PhysRevE.100.042201}.

Mixed-curvature allows for the localisation of scrambling, i.e. quantum information scramble in some regions of the system while remaining coherent in others, a feature not available in fully chaotic or fully regular systems\cite{PhysRevLett.118.086801}. This spatial heterogeneity introduces non-universal scrambling timescales, making the effective scrambling time strongly dependent on the initial state. Consequently, thermal averaging over states can mask this rich, underlying heterogeneity. Taking the logarithm of the OTOCs converts its exponential growth into a straight line, making chaotic behaviour easy to spot. This also helps isolate the time window where genuine Lyapunov growth occurs before saturation or scrambling dynamics set in.

Recent analyses of quantum billiards and finite chaotic systems indicate that apparent exponential windows can be very short or even non-existent, especially when interference effects dominate the early dynamics. For instance, Hashimoto \emph{et al.}\cite{Hashimoto2017} reported that stadium billiards do not display a robust long-lived exponential regime, despite being classically chaotic. Note that, the relation between the exponential growth rate of OTOC and the classical Lyapunov exponent is only valid for systems with a suitable semiclassical limit. Hence, OTOC reflects local instability rather than asymptotic chaos.

\begin{figure*}[hbt!]
	\centering
\subfigure{\includegraphics[width=\linewidth]{"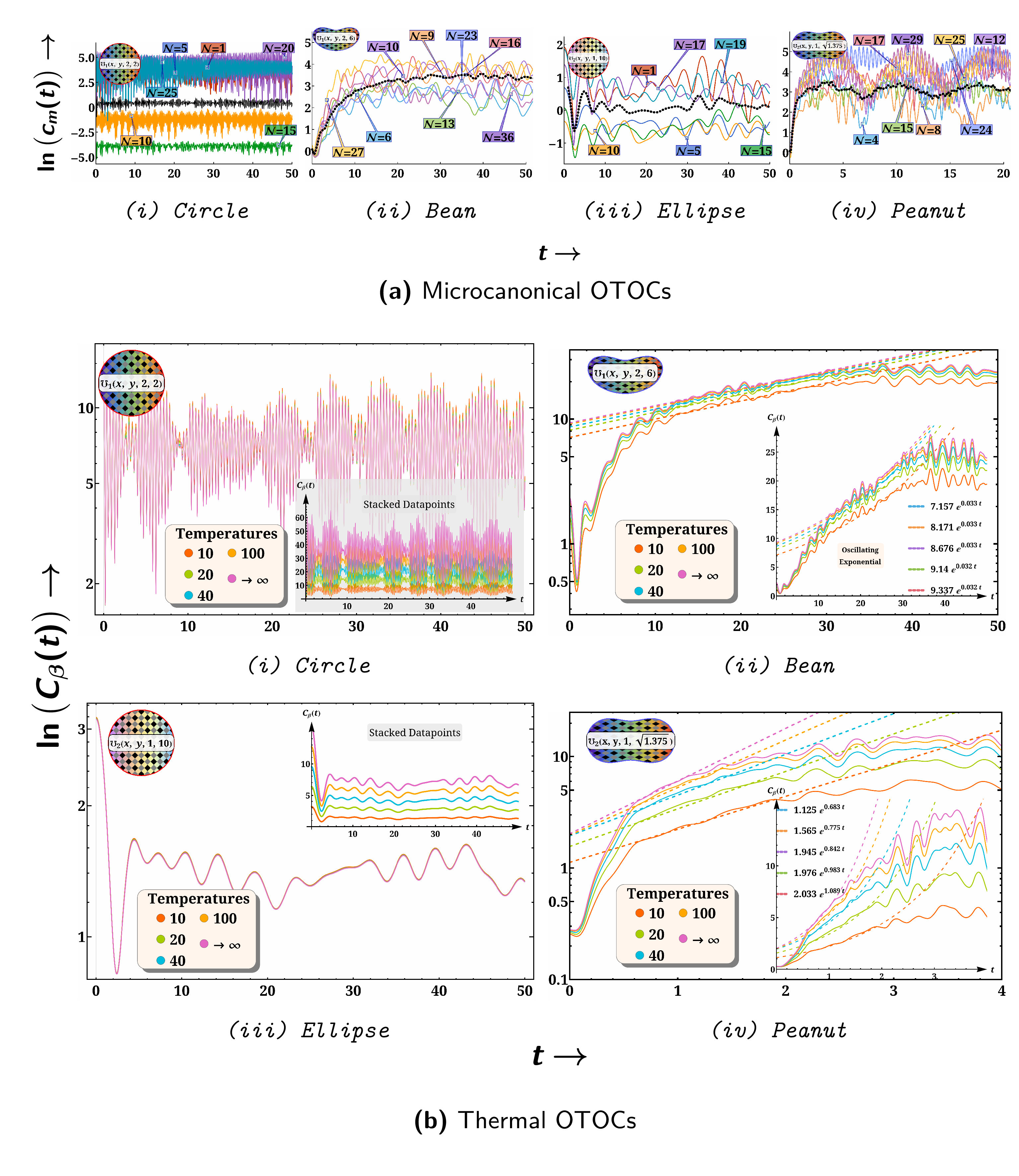"}}
	\caption{\label{fig16} Temporal growth of OTOCs in quantum billiards with distinct geometries. \textbf{Upper panel:} Time evolution of $ \log\qty(c_{m}(t)) $ for individual eigenstates ($\mathcal{N}$). Here, the black dotted lines show the average microcanonical OTOC behaviour. \textbf{Lower panel:} Time evolution of $ \log\qty(C_{\beta}(t)) $. As expected, $ \ln(C_{\beta}(t)) $ exhibits short-time linear growth (or exponential in $C_{\beta}(t)$) for bean and peanut billiards. The dashed fitted lines show the exponential behaviour. Conversely, circular and elliptical billiards show aperiodic random fluctuations, indicating integrability of the systems. While chaotic billiards show temperature sensitivity, integrable billiards remain unaffected. The inset diagrams in (b)(i) \& (b)(iii) show stacked data points at different temperatures to compare trends across different datasets in a cumulative manner, while in (b)(ii) and (b)(iv), the insets show $ C_{\beta}(t) $ vs $ t $ with the extracted exponential-fitting window indicated by a dashed line.}
\end{figure*}

\cref{fig16}(a) (upper panel) presents the time evolution of $ \ln(c_{m}(t)) $ for various geometries and energy levels $ \mathcal{N} $. Here, the black dotted lines indicate the average microcanonical OTOC behaviour. In contrast, the lower panel of \cref{fig16} displays $ \ln(C_{\beta}(t)) $ as a function of time, taking the thermal average of the microcanonical OTOCs. In both panels, the appearance of a straight-line segment signals an exponential growth regime of the OTOC\footnote{Taking the logarithm allows exponential growth to appear as a straight line, making it easier to identify chaotic behaviour.}. $ c_{m}(t) $ for circular and elliptical billiards exhibits persistent oscillations without sustained growth, characteristic of integrable dynamics where operator spreading is limited. Meanwhile, bean- and peanut-shaped billiards display a rapid rise in OTOC followed by saturation, consistent with chaotic dynamics and faster scrambling of quantum information. We have taken thermal averages at different temperatures to improve the distinguishability between chaotic and regular behaviour.

In \cref{fig16}(b-i \& -iii), respectively for circular and elliptical billiard, $ C_{\beta}(t) $ shows random oscillations without discernible exponential growth regime, suggesting the absence of classical-like chaotic behaviour. Here, $ C_{\beta}(t) $ remains relatively stable across all simulated times and temperatures. This lack of temperature dependence implies that thermal fluctuations are not driving operator growth in regular billiards. The inset graphs in the above mentioned figures show stacked data-points at different temperatures\textemdash meaning each temperature series is offset by a fixed vertical shift so that curves do not overlap. This layout is particularly useful for visualising and comparing trends across different datasets in a cumulative manner.

In \cref{fig16}(b-iv), for and peanut-shaped, the presence of rapid growth window ($ C_{\beta}(t)\propto\mathrm{e}^{2 \lambda_{\rm eff} t} $) indicates genuine chaotic dynamics. This behaviour is inherently transient in bounded quantum systems. Furthermore, the saturation behaviour itself serves as a dynamical discriminator: integrable systems exhibit persistent oscillations in the saturation region, while chaotic systems show exact thermalisation to a constant value. In peanut-shaped billiard, the linear (in semi-log plot) segments are visible at early times and have been marked with dashed-fitted-lines. The slopes of these fitted lines increase mildly with temperature but remain bounded. This is fully consistent with theoretical expectations that thermal OTOC growth is constrained\textemdash most famously by the \textit{Maldacena-Shenker-Stanford} (MSS) bound\cite{Maldacena2016} $ \lambda \le 2\pi / \beta $. 

In contrast, the bean-shaped billiard (\cref{fig16}(b-ii)), exhibits a markedly different early-time profile compared to the aperiodic behaviour seen in circular (\cref{fig16}(b-i)) and elliptical (\cref{fig16}(b-iii)) billiards, as well as the sharp, sustained growth observed in the peanut-shaped billiard from (\cref{fig16}(b-iv)). Here, the thermal OTOC develops an oscillatory, ladder-like growth window with only weak temperature dependence. This modulation signals the presence of residual regular structures, where interference between ergodic and partially localised or scarred states occurs at characteristic frequencies determined by their energy separations. As a result, the growth is not purely exponential; rather, oscillations are superimposed on a slowly increasing envelope, consistent with the very small fitted Lyapunov exponent. The origin of this slow-growth behaviour can be traced to the structure of the eigenstates. For low-lying eigenstates (small $\mathcal{N}$) in a billiard, the characteristic length scale $L=\frac{\int \abs{\psi}^{2}\dd{A}}{\int\abs{\grad{\psi}}^{2}\dd{A}}$ over which the probability density $\psi^{2}$ varies is comparable to the linear dimensions of the system itself. For a state with wave-number $k_{\mathcal{N}} ~(=\frac{\sqrt{2m \mathcal{E}}}{\hbar})$, this scale is proportional to the reduced de Broglie wavelength $1/k_{\mathcal{N}}$, i.e., $L\sim\frac{1}{k_{\mathcal{N}}}$ up to a geometry‑dependent constant. Consequently, these long‑wavelength modes cannot resolve the fine details of the boundary, such as local curvature or the presence of corners \cite{PhysRevLett.123.010601, huang2025thirdorderperturbativeotocharmonic, Das2025, Hashimoto2017}. Instead, the eigenstates respond primarily to the global geometry (e.g., the area), rendering distinctions between integrable and weakly chaotic shapes less pronounced. Only when $L$ becomes much smaller than the local radius of curvature do eigenstates begin to resolve boundary intricacies, allowing classical chaotic signatures to emerge in quantum observables. Therefore, in the bean-shaped billiard, the dominance of long-wavelength modes at low $\mathcal{N}$ suppresses strong chaotic growth and produces the observed oscillatory OTOC behaviour with weak exponential amplification.


When comparing bean- and peanut-shaped, OTOC for the later shows steeper initial slopes (larger $ \lambda_{\rm eff} $) than the former. This suggests that the microscopic parameters for the system in peanut open stronger scrambling channels or faster operator complexity growth. This observation agrees with our classical analysis and other independent quantum chaos diagnostics, all of which indicate that the proposed mixed-curvature billiards are predominantly chaotic.

Furthermore, in contrast to integrable billiards, the saturation level of OTOC in bean- and peanut-shaped billiards rises with increasing temperature as shown in \cref{fig16}(b-ii \& -iv), respectively. At low temperatures, the OTOC grows more slowly because thermal occupation is dominated by low-energy states, which often have weaker chaos. As temperature increases, more higher energy states contribute. In strongly chaotic systems, $ \lambda_L $ approaches the classical Lyapunov exponent as $ T \to \infty $. After the exponential growth, the OTOC reaches a plateau (saturation). The height of the plateaus also depends on temperature ($ \propto T $)\cite{Maldacena2016}. These temperature dependency of $ C_{\beta}(t) $ for integrable and chaotic billiards closely mirrors that of $ C_{S}(t) $. In integrable systems, temperature dependency is negligible, whereas in chaotic billiards, it becomes noticeable.


In addition to this, as underpinned by \cite{Chernov1997, Garrido1997, 10.1063/1.5099446}, for a billiard of fixed shape the Lyapunov exponent scales as $ \lambda_{L}\propto\frac{1}{\sqrt{\abs{\Omega}}} $ in connection with the mean free time between collisions.  Since chaotic instability in billiards is generated at boundary collisions, fewer collisions per unit time lead to a smaller Lyapunov exponent. In this sense, enlarging the billiard suppresses dynamical instability. This is a purely geometric suppression of chaos. This geometric scaling has direct implications for the short-time behaviour OTOC. Because early-time OTOC growth reflects the underlying classical sensitivity to initial ICs in the semiclassical regime, a smaller Lyapunov exponent translates into a slower initial growth rate. Consequently, larger billiards are expected to exhibit slower scrambling at short times. This trend aligns with the area ordering in shown in \cref{table4}. The elliptical billiard, having the largest domain among the geometries considered, displays the slowest operator growth, whereas the peanut-shaped geometry, with the smallest area, shows the fastest (\cref{fig16}). The elliptical billiard, having the largest domain among the geometries considered, displays the slowest operator growth, whereas the peanut-shaped geometry shows the fastest. The circle and bean-shaped billiard fits naturally within this ordering. This consistent area dependence reinforces the geometric origin of the scrambling rate.

\subsubsection{\label{sec4Di}OTOC and spatial morphology of eigenstates}

\begin{figure*}[hbtp!]
	\centering
	\includegraphics[width=\linewidth]{"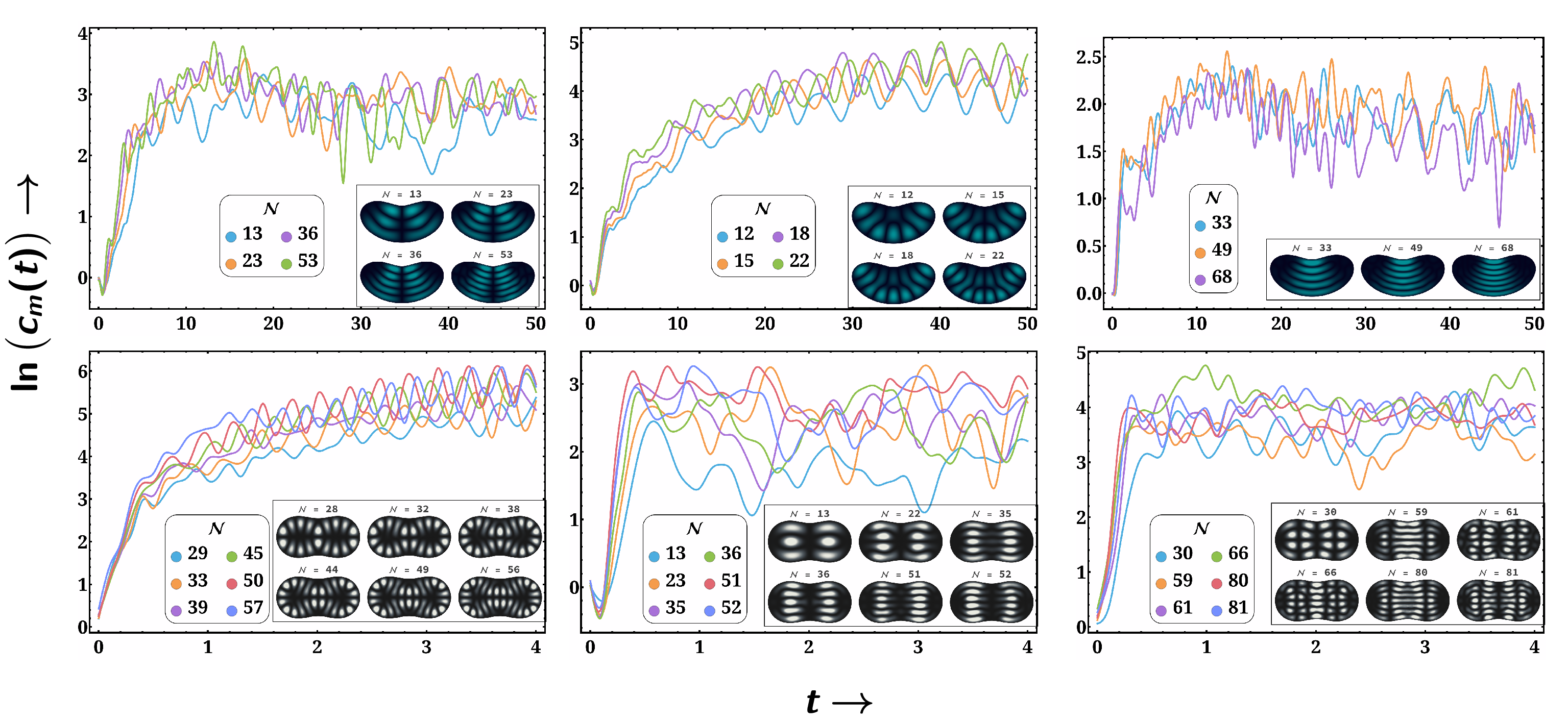"}
	\caption{\label{fig17}Time evolution of the microcanonical OTOC for selected eigenstates plotted together with their corresponding probability-density profiles (insets). \textbf{Upper panel:} Semi-logarithmic plot for bean-shaped billiard. \textbf{Lower panel:} Semi-logarithmic plot for peanut-shaped billiard. Each curve corresponds to a distinct nondegenerate energy eigenstate, while the insets depict the associated probability density distributions. Eigenstates belonging to the same morphology class exhibit indistinguishable scrambling behaviour at early times ($\propto t^{2} $). Eigenstates with visibly different morphology, by contrast, produce distinct OTOC curves, demonstrating the strong correlation between spatial structure of quantum eigenmodes and operator-growth dynamics in chaotic billiards.}
\end{figure*}

Eigenstates with similar probability densities distribution or nearly identical spatial profiles (same nodal/scar structure), tend to share related semiclassical structures, even when their eigenenergies satisfy $ \mathcal{E_{N}}\neq \mathcal{E_{M}} $. As shown in \cref{eq24}, $ c_{m}(t) $ entirely depends on the spectrum $ \mathcal{E_{N}} $ and the position matrix elements $ x_{\mathcal{NM}} $. The value of $ x_{\mathcal{NM}} $ depend sensitively on where the eigenfunction has weight\footnote{The weight of an eigenfunction refers to where the probability density is significantly high or concentrated.}. Two states with similar spatial morphology have almost the same ``support region''\footnote{Support region is the spatial subset where the eigenfunction has appreciable amplitude or structure.} in configuration space; thus the operator probes nearly the same region, generating comparable sets of matrix elements $ x_{\mathcal{NK}} $, but not necessarily identical values. Under unitary evolution, these states scramble information in a parametrically similar way over a characteristic time window. However, they eventually diverge because of differences in energy level spacing and off-diagonal couplings. As a result, visually identical spatial patterns across different eigenstates yield the same dynamical response: identical time-dependence of microcanonical OTOC for local operators.

Because local operator spreading is still microscopic, this regime, near $t=0$, does not diagnose chaos; rather, it encodes the algebraic ``curvature'' of the operator space. Early-time OTOC behaviour is generically quadratic ($\propto t^{2} $) due to the Baker–Campbell–Hausdorff expansion and locality. This quadratic term reflects the local structure of the Hamiltonian. Spatial deformation in curvature of the potential landscape generate spatially dependent deformation of phase-space sensitivity\cite{Das2025}. The later exponential regime, if it exists, grows as $ \propto \mathrm{e}^{2 \lambda_{L} t}$ , but it always emerges from this quadratic seed.

\cref{fig17} shows microcanonical OTOC for several eigenstates of bean- and peanut-shaped billiards whose probability densities are illustrated in the insets. Each coloured curve corresponds to a different eigenstate index $ \mathcal{N} $. A striking observation in this graph is that the OTOC curves do not distribute randomly across the plot; instead, they form morphology-based clusters. Eigenstates that share the same morphological family (nearly identical probability densities) produce OTOC curves that lie almost on top of each other across the two billiards. It is important to emphasise that this similarity extends only over time windows short compared to the $t_{E}$. In contrast, states with different spatial morphology naturally scramble information at different rate and exhibit visibly different OTOC dynamics. The graph thus provides clear evidence that the geometry of the eigenstate, not merely its energy, governs how efficiently an operator spreads information in Hilbert space\cite{Das2025}.

In summary, chaotic eigenstates with matching spatial morphology show very similar sets of position-operator matrix elements and, by extension, OTOC dynamics, despite their varying eigenenergies. Therefore, ``morphology multiplets'' exhibit analogous scrambling within the OTOC. Since OTOC is operator-dependent, another operator (momentum, local projector, or smooth coarse-grained probe) may produce different clustering.

\section{\label{sec5} Conclusion}

In bean- and peanut-shaped billiards, competing focusing and defocusing mechanisms induce orbital destabilisation, characterised by a period of transient confinement preceding chaotic expulsion. In the absence of flat walls in these mixed-curvature systems, curvature-driven chaos dominates in both classical and quantum domains. These models also facilitate chaos control by fine-tuning geometric parameters (curvature adjustment). Therefore, the nature of dynamics depends on the curvature of the confining boundary.

In summary, our study shows that bean- and peanut-shaped billiards, having smooth focusing (concave) and defocusing (convex) walls, produce dominant chaotic dynamics for a majority of ICs. In the bean-shaped billiard, having lower symmetry, there are periodic and quasi-periodic trajectories  for a subset of ICs. These appear as isolated points (periodic orbit) and island structures (quasi-periodic orbit) within scattered points (scattered points) in the Poincaré section. In contrast, the more symmetric peanut-shaped billiard supports quasi-periodic motion for only a few ICs with no periodic motion. However, a few special ICs produce slowly diverging, nearly periodic trajectories, which are linked to the formation of quantum scars. Increased scarring in the bean and peanut-shaped billiards establishes a clear visual connection among geometric symmetry, classical periodic orbits, and quantum eigenstate localisation, reinforcing the quantum-classical correspondence in chaotic systems. Furthermore, the classical Lyapunov exponent remains positive for these billiards, quantitatively confirming the underlying sensitivity to ICs and the predominance of chaotic dynamics.

These classical dynamics manifest themselves in the quantum regime. We observe both scars and super-scars in both models, with the peanut producing more scarred states than the bean. Statistical diagnostics such as the NNSD, CLSD, LSR, and the spectral staircase function collectively reveal a dominant chaotic behaviour for our mixed-curvature billiards. Alongside these conventional diagnostic methods, we have also employed two relatively novel tools; SC and OTOCs.

Spectral complexity acts as a dynamical seismograph of quantum chaos, progressing through quadratic, linear, and logarithmic regimes that expose how correlations grow and universality emerges. Chaotic systems saturate rapidly, typically near the Heisenberg time, reflecting their strong mixing and universal spectral correlations. Integrable systems, with their Poissonian level statistics and weak mixing, saturate only on extremely long time scales or effectively never; making the saturation time a clear discriminator between order and chaos. Despite their rapid and slow transitions from quadratic to linear growth regimes, both bean- and peanut-shaped billiards saturate near the Heisenberg time, confirming their chaotic nature, while integrable billiards like the circle or ellipse saturate only at much later times.

OTOC shows initial growth consistent with chaos, though the exponential regime is short-lived and modest in amplitude. This reflects the mixed nature of the dynamics, where chaos is dominant but tempered by residual regular structures. Integrable billiards, by contrast, show irregular, non-exponential fluctuations with no well-defined growth phase. The scrambling timescale scales with the billiard's area, with larger domains producing slower OTOC onset. Complementing this, eigenstates sharing similar probability-density morphologies yield nearly identical microcanonical OTOC behaviour. Specifically, in the peanut billiards, the OTOCs exhibit an exponential-to-saturation profile, which is characteristic of chaos. In the bean-shaped billiard, the growth rate is much slower with an oscillatory profile, highlighting that information scrambling is strongly governed by the characteristic length scale of low-energy eigenstates, thereby limiting sensitivity to boundary-induced chaos. Additionally, integrable systems (circle and ellipse) show minimal temperature dependence in their OTOCs, whereas chaotic billiards (bean and peanut) display strong sensitivity to temperature, further reinforcing the distinction between regular and chaotic dynamics. Note that OTOC's exponential growth is a transient semiclassical signature, not an asymptotic marker of chaos. 

To conclude, SC and OTOCs do not replace traditional spectral diagnostics; they extend them by uncovering temporal, geometric, and scrambling-related features that remain invisible to conventional measures. Together, they provide a richer and more discriminating picture of chaos in quantum billiards. The presence of quantum chaos as revealed by these quantum mechanical tools (traditional and newer) corresponds strongly to the dominant classical chaotic behaviour.

\begin{acknowledgments}
The authors extend their gratitude to Subhash C. Mahapatra and Soudamini Sahoo for their generous provision of computational resources, which greatly aided in different portions of the numerical computations. The authors also acknowledge the use of AI tools for editorial assistance in improving manuscript clarity.
\end{acknowledgments}

\section*{Author Contributions}
P.P.D. and B.G. conceptualised the work; P.P.D. and T.P. developed the methodology; P.P.D. and T.P. implemented the software, curated the data, and prepared visualisations; P.P.D. performed the formal analysis and drafted the manuscript; T.P. and B.G. validated the results, reviewed, and edited the manuscript; B.G. supervised the project, and administered the work.

\appendix
\section{\label{apndx-A}Visualisation of quantum scars}

\begin{figure}[H]
	\centering
	\includegraphics[width=\linewidth]{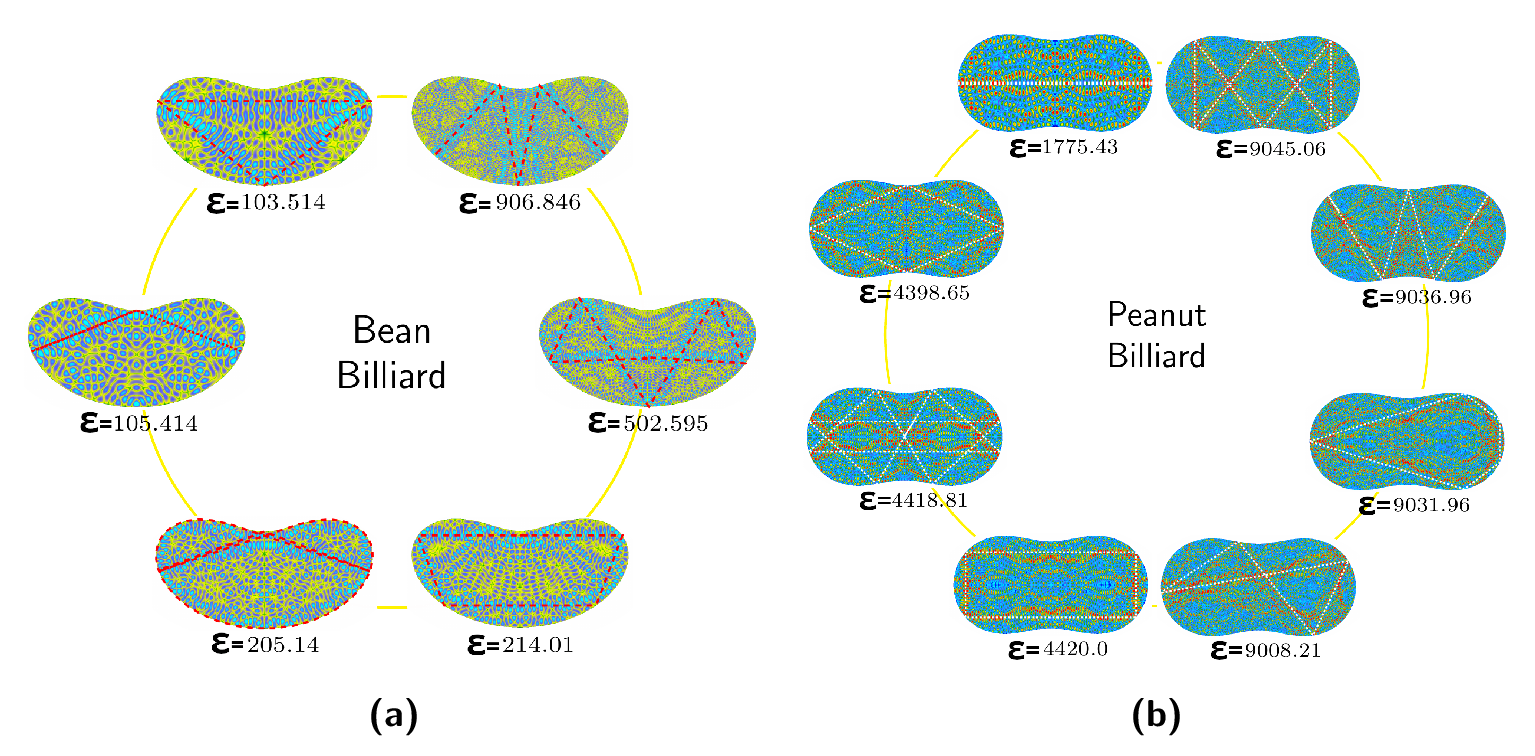}
	\caption{Quantum scars for slow diverging trajectories in (a) bean-shaped billiard and (b) peanut-shaped billiard. The red and white dashed line represent slow diverging trajectories.}
	\label{fig18}
\end{figure}

In \cref{fig18}, probability densities of some of these eigenstates. These eigenstates are subtlety scarred by the slow diverging trajectories, and thus weakly breaking the ergodicity. These initially formed, high-density regions persist, remaining visible indefinitely.

\section{\label{apndx-B}Spectral Complexity as $ \beta\rightarrow0 $}

As $ T \to \infty $ or $ \beta \to 0 $, the Boltzmann weights become uniform, i.e. every eigenstate contributes equally. Since for non-degenerate eigenvalue spectra, $ Z(2 \beta)\to\mathcal{N }=D $ as $ \beta\to 0 $ and $ (\mathbf{\mathrm{e}}^{-\beta(\mathcal{E}_{\mathcal{M}}+\mathcal{E}_{\mathcal{N}})})\to 1 $. As a result, \cref{eq21} becomes
\begin{eqnarray}\label{eq25}
	C_{s}(t)&\xrightarrow{\beta\to 0}& \dfrac{1}{D^{2}} \sum_{\mathcal{M} \neq \mathcal{N}}^{D} \Bigg(\dfrac{\sin(\dfrac{t~\Delta \mathcal{E}_{\mathcal{MN}}}{2})}{\dfrac{\Delta \mathcal{E}_{\mathcal{MN}}}{2}}\Bigg)^{2}\\
	&\leq& \dfrac{1}{D^{2}} \sum_{\mathcal{M} \neq \mathcal{N}}^{D} \Bigg(\dfrac{2}{\Delta \mathcal{E}_{\mathcal{MN}}}\Bigg)^{2}
\end{eqnarray}
where, $ \Delta\mathcal{E}_{\mathcal{MN}}= \mathcal{E}_{\mathcal{M}}-\mathcal{E}_{\mathcal{N}} $. Thus the smallest $ \abs{\Delta\mathcal{E}_{\mathcal{MN}}} $ contribute dominantly to the saturation value of $ C_{s} (t) $.

So at infinite temperature, SC becomes maximally sensitive to global spectral correlations and the system's full Hilbert-space structure. This makes $ C_{s}(t) $ a global probe of the entire spectrum rather than a local (in energy) probe. This is the usual ``infinite temperature'' limit used to study overall spectral correlations and universal RMT behaviour\cite{Cotler2017}. However, at finite non-zero temperature, SC probes within a ``Boltzmann window'' of width $ \sim \frac{1}{\beta} $ in energy\cite{PhysRevE.111.014135}. Thus, at finite $ \beta $, SC gives spectral correlations localised around energies $ \mathcal{E}\leq \order{\frac{1}{\beta}} $, or more generally within the spectral region where the weight $ \mathbf{\mathrm{e}}^{-\beta \mathcal{E}} $ is non-negligible. As a result, SC captures the local spectral features. 

So by choosing $ \beta $ we control how large an energy band contributes: smaller $ \beta $ (higher temperature) broadens the band; larger $ \beta $ (lower temperature) narrows the band, thus focusing on spectral statistics near some lower-energy region.

\bibliography{Bibliography}

\end{document}